\documentclass{article}
\usepackage{forest}
\usepackage{setspace}
\usepackage{biblatex} 
\usepackage{amsmath}
\RequirePackage{fullpage}
\RequirePackage[most]{tcolorbox}
\RequirePackage{chngcntr}
\RequirePackage{amsmath, amsthm, amsfonts, amssymb}
\RequirePackage{graphicx}
\RequirePackage{comment}
\RequirePackage{subcaption}
\RequirePackage{caption}
\RequirePackage{enumitem}
\RequirePackage{physics}
\RequirePackage{tikzlings}
\RequirePackage{comment}
\RequirePackage{tikz}
\usetikzlibrary{tikzmark}
\usetikzlibrary{calc, karnaugh, arrows, circuits.logic.US, angles, quotes,fit,positioning}
\usetikzlibrary{quantikz}
\usepackage{kvmap}
\usepackage{pstricks,multido}
\psset{dimen = m, fillstyle = solid}
\usepackage{xcolor}
\usepackage[margin=1in]{geometry}
\usepackage{blindtext}
\usepackage{subfiles} 
\usepackage{ragged2e}
\usepackage{changepage}
\usepackage{blochsphere}
\usepackage{mathtools}
\usetikzlibrary{arrows.meta}
\usepackage{float}

\title{\vspace{-1.6cm}EQB: Synthesizing Permutative Quantum  Gates and Circuits Using Rotation-Based Group Decomposition}

\newcommand{\CellWithForceBreak}[2][c]{
\begin{tabular}[#1]{@{}c@{}}#2\end{tabular}}

\begin{document}
\date{\vspace{-10ex}}
 
\maketitle
    \begin{center}
        Ishani Agarwal
        
        ishani2807@gmail.com

        orcid.org/0009-0009-1299-929X
    \end{center}
    
    \begin{center}
        Miroslav Saraivanov

        michaelsaraivanov@yahoo.com
        
    \end{center}

    \begin{center}
        Ali Al-Bayaty

        albayaty@pdx.edu

        orcid.org/0000-0003-2719-0759

    \end{center}
     
    \begin{center}
        Marek Perkowski

        h8mp@pdx.edu

        orcid.org/0000-0002-0358-1176
        
    \end{center}

    \begin{center}
        Department of Electrical and Computer Engineering, Portland State University 
        
        Portland, Oregon 97201, United States
    \end{center}
\vspace{-2ex}

\[\textbf{Abstract}\]
\begin{adjustwidth}{2.3cm}{2.3cm}
\justify{The decomposition from the group theory-based methods of Sasao and Saraivanov is extended to design binary quantum cascades, using the quantum rotational gates by the X-axis (CNOT and RX), Y-axis (RY), and Z-axis (controlled-Z) of the Bloch sphere. A class of local transformations is also presented to simplify the final canonical cascade circuits. Our proposed methodology is well suited for quantum layouts, as each single-qubit gate has one target qubit and each double-qubit gate has one control qubit and one target qubit, thereby never creating a graph of triangular connectivity.}
\end{adjustwidth}

\[\textbf{Keywords}\]
\begin{adjustwidth}{2.3cm}{2.3cm}
\justify{Permutative quantum gates, group theory, group decomposition,  binary quantum cascades, quantum layouts, Bloch sphere}
\end{adjustwidth}

\section{Introduction}

\indent 




There are existing methodologies to relate group logic to the synthesis of cascade circuits, as presented in [\ref{0}-\ref{7}]. However, in this paper, we propose a new methodology to synthesize functions using quantum rotational gates. In [\ref{2}], a group function decomposition is derived based on Shannon's decomposition. We utilized such a decomposition method to create rotation-based quantum circuits. The only quantum gates we used in this method are based on the quantum rotations around the axes of the Bloch sphere [\ref{8}], which are the X-axis (CNOT and RX gates), Y-axis (RY gates), and Z-axis (controlled-Z or CZ gates). This method allows us to choose either RX or RY gates, to create a more general and adaptable approach for different quantum layouts and technologies. For recent quantum technologies, the hardware rotation precision for RX and RY gates is around $\pi \cdot 10^{-3}$, as stated by IonQ [\ref{9}].

The quantum circuits that are realized using our proposed method also prioritize the quantum layout, since our proposed quantum gates always have a maximum of one control qubit and one target qubit. This is advantageous for the layouts in some modern quantum technologies since a single qubit has four neighbored qubits (at most) that are arranged in square lattices. Having only one control qubit and one target qubit allows a quantum circuit to be mapped into such quantum layouts. The quantum layout problem was discussed in [\ref{10}-\ref{13}], but it is not further discussed in this paper. Our proposed method only relates to the canonical cascades for quantum gates and circuits.

This paper is organized into the following sections. Section 2 provides background information on the Bloch sphere, group theory, and a previous methodology that uses group decomposition to synthesize quantum circuits. Section 3 introduces our proposed exact quantum binary (EQB) methodology with some examples. Section 4 lists the local transformations that are used in our EQB methodology. Section 5 presents additional examples of our EQB methodology, including multi-valued functions. Section 6 presents our EQB methodology as the ``EQB synthesis tool," with comparisons to different benchmark functions. Appendix A provides an in-depth proof of the formula for the Walsh Spectrum of a function in terms of its truth vector. Appendix B discusses the multi-input group decomposition expressions for the intermediate cascades and decompositions of multi-variable functions.



\section{Background}

This section covers the topics related to the visualizations of the Bloch sphere, group theory, and how to synthesize binary inputs multi-valued outputs quantum circuits using group decomposition with examples.

\subsection{Bloch Sphere}
\begin{figure}[!ht]
    \centering
    \scalebox{.8}{
    \begin{tikzpicture}[line cap=round, line join=round, >=Triangle]
  \clip(-2.19,-2.49) rectangle (2.66,2.58);
  \draw [shift={(0,0)}, lightgray, fill, fill opacity=0.1] (0,0) -- (56.7:0.4) arc (56.7:90.:0.4) -- cycle;
  \draw [shift={(0,0)}, lightgray, fill, fill opacity=0.1] (0,0) -- (-135.7:0.4) arc (-135.7:-33.2:0.4) -- cycle;
  \draw(0,0) circle (2cm);
  \draw [rotate around={0.:(0.,0.)},dash pattern=on 3pt off 3pt] (0,0) ellipse (2cm and 0.9cm);
  \draw (0,0)-- (0.70,1.07);
  \draw [->] (0,0) -- (0,2.5);
  \draw [->] (0,0) -- (-0.81,-0.79);
  \draw [->] (0,0) -- (2.2,0);
  \draw [dotted] (0.7,1)-- (0.7,-0.46);
  \draw [dotted] (0,0)-- (0.7,-0.46);
  \draw (-0.08,-0.3) node[anchor=north west] {$\varphi$};
  \draw (0.01,0.9) node[anchor=north west] {$\theta$};
  \draw (-1.01,-0.72) node[anchor=north west] {$\mathbf {{x}}$};
  \draw (2.09,0.28) node[anchor=north west] {$\mathbf {{y}}$};
  \draw (0,2.6) node[anchor=north west] {$\mathbf {{z}}$};
 \draw (0,2) node[circle, fill, inner sep=1.2] {}; 
  \draw (-0.65,2.5) node[anchor=north west] {$ {|0\rangle}$};
   \draw (0,-2) node[circle, fill, inner sep=1.2] {}; 
  \draw (-0.3,-2) node[anchor=north west] {$ |1\rangle$};
  \draw (0.4,1.65) node[anchor=north west] {$|\psi\rangle$};
  \scriptsize
  \draw [fill] (0,0) circle (1.5pt);
  \draw [fill] (0.7,1.1) circle (0.5pt);
\end{tikzpicture}}
\caption{The Bloch sphere and its three axes (X-axis, Y-axis, and Z-axis), for visualizing the quantum state $|\psi\rangle$ of a single qubit rotating around the Y-axis and Z-axis by the angles $\theta$ and $\varphi$, respectively.}
  \label{BlochSphere}
\end{figure}
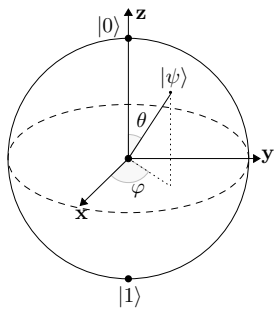

The Bloch sphere geometrically visualizes the quantum states of a single qubit, as shown in Figure \ref{BlochSphere}.  The quantum state of a qubit can be represented along the three axes (X-axis, Y-axis, and Z-axis) of the Bloch sphere, e.g., the basis state of $|0\rangle$ or $|1\rangle$ along the Z-axis, the superposition of basis states $(|0\rangle \pm |1\rangle)$ along the X-axis, or the complex superposition of basis states $(|0\rangle \pm i|1\rangle)$ along the Y-axis [\ref{8}]. Note that any two diametrically opposite points on the Bloch sphere are orthogonal.

\begin{figure}[!ht]
    \centering
    \includegraphics[width=3.2cm]{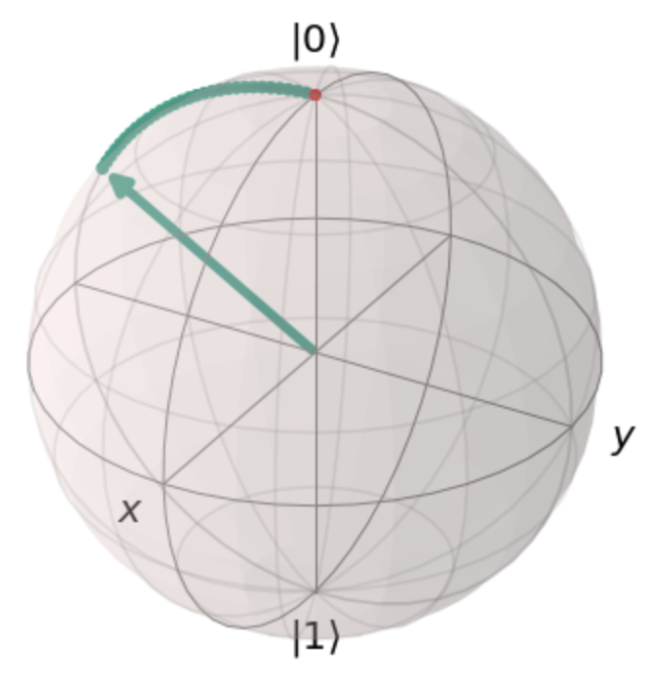}
    \caption{The obtained vector after counterclockwise rotating a $|0\rangle$ state by $\frac{\pi}{3}$ (in radians) around the X-axis of the Bloch sphere.}
    \label{1-60deg}
\end{figure}

An important use of the Bloch sphere is that the progression of a qubit's state can be easily described using the angular rotations around any axis of the Bloch Sphere. For instance, if a qubit in the state of $|0\rangle$ rotates counterclockwise around the X-axis by $\frac{\pi}{3}$ in radians, then its corresponding vector can be easily obtained as illustrated in Figure \ref{1-60deg}.

Rotating the state of a qubit (as a vector) around the X-axis of the Bloch sphere is equivalent to multiplying such a vector by $R_x(\theta)$, where
\[R_x(\theta) = \begin{bmatrix} \cos{\frac{\theta}{2}} & -i\sin{\frac{\theta}{2}} \\ -i\sin{\frac{\theta}{2}} & \cos{\frac{\theta}{2}}   
\end{bmatrix}\text{.}\]

Note that both $R_y(\theta)$ and $R_z(\theta)$ similarly follow the above definition, where
\[R_y(\theta) = \begin{bmatrix} \cos{\frac{\theta}{2}} & -\sin{\frac{\theta}{2}} \\ \sin{\frac{\theta}{2}} & \cos{\frac{\theta}{2}}   
\end{bmatrix}\text{ and }R_z(\theta) = \begin{bmatrix} e^{-i\frac{\theta}{2}} & 0 \\ 0 & e^{i\frac{\theta}{2}}
\end{bmatrix}\text{.}\]

\subsection{Group Theory}
A binary operation performed on two elements in a set obtains the third element of this set. A group, which is denoted as $\langle G, \star \rangle$, is an algebraic structure consisting of a set ($G$) with a binary operation ($\star$), to satisfy the four properties (closure, identity, associativity, and invertibility) as follows.

\begin{enumerate}
    \item Closure: $a \star b \in G$; $\forall a,b \in G$.
    \item Identity: There exists a unique $e \in G$, such that $e \star a = a \star e = a$; $\forall a \in G$.
    \item Associativity: $a \star (b \star c) = (a \star b) \star c$; $\forall a,b,c \in G$.
    \item Invertibility: $\forall a \in G$, there exists $b \in G$, such that $a \star b = b \star a = e$ where $e$ is the identity element.
\end{enumerate}

\indent The order of a group is equal to the number of unique elements in its set. An abelian group, $\langle G, \star \rangle$, is a commutative group that satisfies $g_1 \star g_2 = g_2 \star g_1$, $\forall g_1, g_2 \in G$. A cyclic group is a group that is generated by one of its elements, $g$. For instance, the group $\langle g^0, g^1, g^2, g^3 \rangle$ is a cyclic group of order 4, when $g^4 = g^0$. The dihedral group, which is denoted as $D_n$, is a non-abelian group of order $2n$ that consists of the rotational and reflectional symmetries of a regular $n$-gon. In general, $D_n =  \langle a^0, a^1 \ldots a^{n-1}, g, a^1g, a^2g \ldots a^{n-1}g \rangle$, where $a^n = a^0 = I$ and $g^2 = g^0$, and $D_n = C_n \times C_2$, where $C_k$ is the cyclic group of order $k$ and $k \geq 1$.

\indent An infinite group is a group that has an infinite number of elements [\ref{14}]. For instance, the group of integers with the operations of addition is an infinite group. An infinite dihedral group is an infinite group that has: (i) properties analogous to those of finite dihedral groups, and (ii) the representation $\langle g, a | g^2 = I, gag = a^{-1} \rangle$ = $\langle I, a^1, a^2 \ldots a^{t}, g, a^1g, a^2g \ldots a^tg \rangle$, where $|$ means $such~that$, and a finite $t$ does not exist, such that $a^t = I$.

\subsection{Synthesis of Binary Input Multi-valued Output Quantum Circuits Using Group Decomposition}

\indent As presented in [\ref{0}, \ref{2}], binary input multi-valued output functions can be synthesized using group function decomposition and Walsh Spectrum [\ref{1}].

\begin{figure}[!ht]
    \centering
    \scalebox{.85}{
    \includegraphics[width=6.5cm]{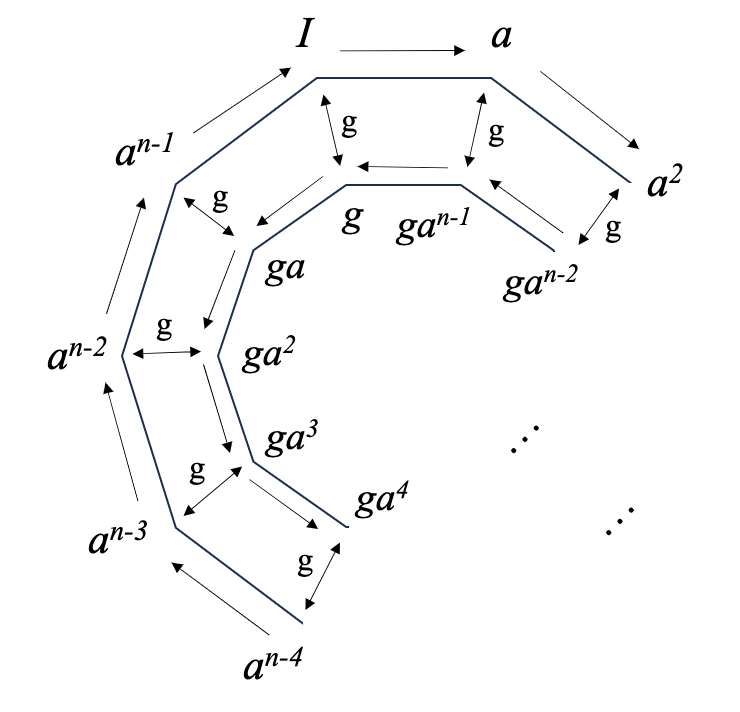}}
    \caption{The group map of a dihedral group ($D_n$).}
    \label{Dn}
\end{figure}

First, a dihedral group ($D_n$) is generated. Group $D_n$ for arbitrary prime number $n$ can be generated using the elements $a$ and $g$, such that $a$ permutes the $n$th rail to the ($n+1$)th rail, where the last rail mapped to the top rail, and $g$ swaps the $i$th rail with the ($n-i$)th rail, $\forall i \neq 0$. Figure \ref{Dn} illustrates the group map of $D_n$.

Second, we derive an expression to determine the group decomposition expression of a function. From [\ref{1}, \ref{15}],
\begin{equation}F(\hat{X},x_n) = F_a(\hat{X})g^{x_n}F_b(\hat{X})g^{x_n} = a^{f_a(\hat{X})}g^{x_n}a^{f_b(\hat{X})}g^{x_n}\text{,}\end{equation} 
which is an expansion similar to Shannon's Expansion [\ref{16}], where $x_n$ is a two-valued input variable and $F_a(\hat{X})$ and $F_b(\hat{X})$ denote group functions that do not depend on $x_n$ with $\hat{X}=(x_1, x_2, \ldots, x_{m-1}) \in X^{m-1}$. Using  Eq. (1), \begin{equation}\vec{w} = (W_1^{-1}\vec{F})\end{equation} is the derived, where $\vec{F}$ is the truth vector of a function, $W_1$ is the first-order Walsh Matrix as stated in Eq. (3), and $\vec{w}$ is the Walsh Spectrum of $\vec{F}$. In general we let $W_n$ denotes the $n$th-order Walsh Matrix where $W_{n+1} = W_1 \otimes W_n$. A more detailed proof can be found in Appendix A.

\begin{equation}
    W_1 = \begin{bmatrix}
1 & 1  \\
1 & -1 
\end{bmatrix}
\end{equation}

Eq. (2) can be extended to $n+1$ input variables using \[F(x_1,x_2, \ldots x_n, x_{n+1}) = F_a(x_2, \ldots x_{n+1})g^{x_1}F_b(x_2, \ldots x_{n+1})g^{x_1}\text{.}\] A more detailed proof can be found in Appendix B.

\noindent \textbf{Example 1:} Consider the function $f(x_1,x_2) = x_1 \oplus x_2$ as a simple binary XOR function. We will use $D_3$ to create the corresponding cascades. The truth vector for this function is $\vec{F} = \begin{bmatrix} 0 & 1 & 1 & 0\end{bmatrix}^T$. Hence, from Eq. (2), we obtain \[\vec{w} = (W_2^{-1})\vec{F} = (W_2^{-1})\begin{bmatrix}
0 \\ 1 \\ 1 \\ 0 \end{bmatrix}\text{.}\]

where $W_2$ denotes the second-order Walsh Matrix. Next, we compute the inverse of the second-order Walsh Matrix $(W_2^{-1})$, which is the Kronecker product of two first-order Walsh matrices ($W_1\otimes W_1$), as follows.

\[(W_2)^2 = \begin{bmatrix}
4 & 0 & 0 & 0 \\
0 & 4 & 0 & 0 \\
0 & 0 & 4 & 0 \\
0 & 0 & 0 & 4 
\end{bmatrix} = 4\begin{bmatrix}
1 & 0 & 0 & 0 \\
0 & 1 & 0 & 0 \\
0 & 0 & 1 & 0 \\
0 & 0 & 0 & 1 
\end{bmatrix} = 4I_2 = 4W_2W_2^{-1}\]
\[\Rightarrow W_2^{-1} \equiv W_2  \text{ (mod 3)}\]

Therefore, \[\vec{w} = (W_2^{-1})\vec{F} = W_2\begin{bmatrix}
0 \\ 1 \\ 1 \\ 0 \end{bmatrix} = \begin{bmatrix}
2 \\ 0 \\ 0 \\ -2 \end{bmatrix} \equiv \begin{bmatrix}
-1 \\ 0 \\ 0 \\1 \end{bmatrix} \text{ (mod 3)}\text{,}\]

\[\vec{w} = \begin{bmatrix}
    w_a & w_b & w_c & w_d
\end{bmatrix}^T = \begin{bmatrix}
    -1 & 0 & 0 & 1
\end{bmatrix}^T \text{ (mod 3)}\text{.}\]

Note that the group decomposition expression for two variables in the canonical form is
\[a^{f(x_1,x_2)} = a^{w_a}g^{x_2}a^{w_b}g^{x_2+x_1}a^{w_c}g^{x_2}a^{w_d}g^{x_2 + x_1}\text{.}\]

\begin{figure}[!ht]
    \centering
    \includegraphics[width=7cm]{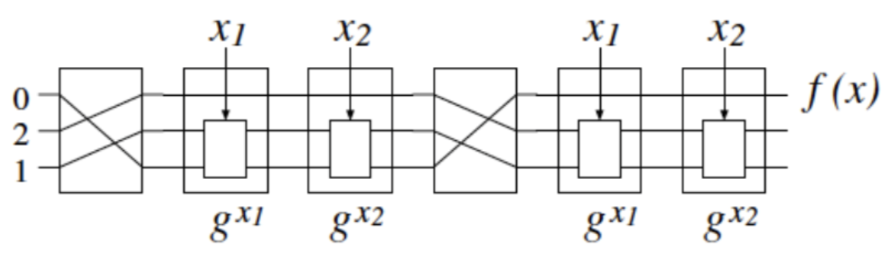}
    \caption{The canonical cascade for $f(x_1,x_2) = x_1 \oplus x_2$.}
    \label{xor-ex}
\end{figure}

Subsequently, by replacing the exponents ($w_a, w_b, w_c, \text{ and } w_d$) with the values obtained from $\vec{w}$ above, the canonical cascade function becomes
\[a^{f(x_1,x_2)} = a^{-1}g^{x_2}a^{0}g^{x_2+x_1}a^{0}g^{x_2}a^{1}g^{x_2 + x_1} = a^{-1}g^{x_1+x_2}a^1g^{x_1+x_2}\text{,}\]
and Figure \ref{xor-ex} illustrates the corresponding canonical cascade for the function $f(x_1,x_2) = x_1 \oplus x_2$.

\section{Methods}

In this section, we propose different methodologies for generating exact quantum binary (EQB) canonical cascades, locally transforming a quantum circuit (from a canonical cascade) for a reduced number of quantum gates, illustrating different examples of reduced quantum circuits, and finally explaining our developed EQB tool for synthesizing a reduced quantum circuit and how to map it into a quantum layout.

\subsection{EQB Canonical Cascades}

\indent In this paper, our proposed method builds on the idea of synthesizing quantum circuits using group decomposition [\ref{1}], and it then generates quantum canonical cascades with outputs that are purely in one of the basis states of $|0\rangle$ or $|1\rangle$. As described in [\ref{0},\ref{2}], binary input multi-valued output functions can be synthesized using group function decomposition and Walsh Spectrum. The quantum circuits in [\ref{0},\ref{2}] were built using permutative gates, such as SWAP, controlled-SWAP (Fredkin), and CNOT gates. However, in this paper, we expand such a methodology using group function decomposition, to synthesize functions using the binary quantum rotational gates around the X-axis (CNOT and RX), Y-axis (RY), and Z-axis (CZ) of the Bloch sphere.

In [\ref{0} - \ref{2}], finite dihedral groups were generated using the elements $a$ and $g$, in which they could be represented as a series of SWAP gates, and the decomposition of arbitrary binary input multi-valued output functions was derived in terms of the group elements. The canonical cascade of a binary input function was represented using elements $a$, which cyclically shifted the wires, and $g$ that are swapped the $i$th wire with the $(k-i)$th wire for $2 \leq i \leq \frac{k+1}{2}$, where $k$ is the number of wires. The exponents of $a$ elements took on the values of the inverse Walsh spectrum with no modulo applied, or $2^{-n}W_n\vec{F}$, where $n$ is the number of input variables, $W_n$ is the Walsh matrix for $n$ variables, and $\vec{F}$ is the function truth vector element.

In this paper, we use the same decomposition derived in [\ref{1}]; however, we use the infinite dihedral group whose elements can be represented as rotations around the axes of the Bloch sphere. Note that a direct method for converting $a^n$ elements from the finite dihedral groups in [\ref{0}, \ref{2}], into elements of the infinite dihedral group is achieved by treating each element as a primitive rotation about any axis (X or Y) of the Bloch sphere, in which it will directly affect the quantum observable state. The exponent $n$ of the element $a^n$ represents a fraction of a rotation between $0^{\circ}$ and $180^{\circ}$. Such that, this exponent can be directly multiplied by $\pi$, to determine the angle of the exact quantum vector rotation. Consequently, the $g^{x_n}$ element creates a reflection axis that effectively conjugates an element to its group inverse [\ref{0}, \ref{2}]. A similar type of conjugation occurs when a qubit rotates around the Z-axis of the Bloch sphere by $180^\circ$, i.e., to form complex conjugate pairs around the Z-axis. Hence, every $g^{x_n}$ element can be converted into a controlled-Z (CZ) gate. Note that regardless of how many input variables we have, all $a$ elements can be converted to $X$ or $Y$ rotations, and all $g$ elements can be converted to $CZ$ gates.


\bigskip

\noindent \textbf{Example 2:} Consider a two-input XOR function, with its truth vector equals $[0,1,1,0]^T$.
\[2^{-n}W_n\vec{F} = 2^{-2}W_2[0,1,1,0]^T = \frac{1}{4} \begin{bmatrix}
    1 & 1 & 1 & 1 \\
    1 & -1 & 1 & -1 \\
    1 & 1 & -1 & -1 \\
    1 & -1 & -1 & 1 
\end{bmatrix} \begin{bmatrix}
    0 \\ 1 \\ 1 \\ 0
\end{bmatrix} =\frac{1}{4}[2,0,0, -2]^T = [\frac{1}{2},0,0, -\frac{1}{2}]^T\]

The canonical cascade for the two input variables equals $((a^{w_a}g^{x_2}a^{w_b}g^{x_2})g^{x_1})((a^{w_c}g^{x_2}a^{w_d}g^{x_2})g^{x_1})$, where $w_a, w_b, w_c$ and $w_d$ are the elements of the Walsh spectrum of this function, and the XOR canonical cascade then becomes: \[a^{\frac{1}{2}}g^{x_2}a^{0}g^{x_2}g^{x_1}a^{0}g^{x_2}a^{-\frac{1}{2}}g^{x_2}g^{x_1}\text{,}\] where the $g^{x_n}$ elements act as CZ gates that provide rotations around the Z-axis of the Bloch sphere, and the $a^{k}$ elements act as RX gates that provide rotations around the X-axis of the Bloch sphere. By combining consecutive $a$ and $g$ elements and using $g^2 = I$, we obtain the XOR canonical cascade of $a^{\frac{1}{2}}g^{x_1}g^{x_2}a^{-\frac{1}{2}}g^{x_2}g^{x_1}$. Note that the last two $g$ elements can always be removed, as rotating the state of $|0\rangle$ or $|1\rangle$ by the Z-axis does not reflect any change. Hence, the canonical cascade of the two-input XOR function is reduced to $a^{\frac{1}{2}}g^{x_1}g^{x_2}a^{-\frac{1}{2}}$.

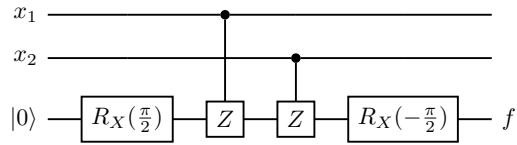
\begin{figure}[!ht]
    \centering
    \scalebox{0.9}{
    \begin{quantikz}
    & \lstick{$x_1$} & \qw & \ctrl{2} & \qw & \qw & \qw \\
    & \lstick{$x_2$} & \qw & \qw & \ctrl{1} & \qw & \qw \\
    & \lstick{$|0\rangle$} & \gate{R_X(\frac{\pi}{2})} & \gate{Z} & \gate{Z} & \gate{R_X(-\frac{\pi}{2})} & \qw \rstick{$f$}
    \end{quantikz}}
    \caption{Intermediate quantum circuit for the two-input XOR function, where $x_1$ and $ x_2$ are the input qubits, and $f$ is the final output qubit of a state $|0\rangle$ (as a non-solution) or $|1\rangle$ (as a solution).}
    \label{XORcircuit1}
\end{figure}

By converting the $a$ elements to RX gates and the $g$ elements to CZ gates, the quantum circuit of $(R_x\frac{\pi}{2})(CZ_1)(CZ_2)(R_x-\frac{\pi}{2})$ is then obtained as depicted in Figure \ref{XORcircuit1}, where $CZ_i$ is a Z gate controlled by the qubit $x_i$. Note that, in Figure \ref{XORcircuit1}, the first two input qubits ($x_1$ and $x_2$) are termed the $control~qubits$, while the last output qubit is termed the $target~qubit$, which is initially set to the state of $|0\rangle$.

\begin{figure}[!ht]
    \centering
    \begin{subfigure}{.4\textwidth}
    \includegraphics[width=3cm]{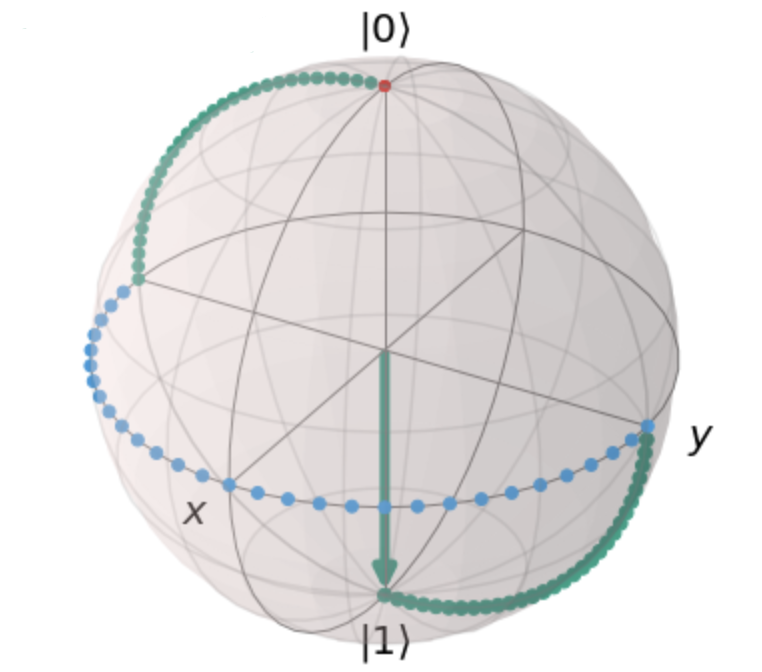}
    \caption{$(R_x\frac{\pi}{2})(CZ_2)(R_x-\frac{\pi}{2}) = 1$}
    \end{subfigure}
    \begin{subfigure}{.4\textwidth}
    \centering
    \includegraphics[width=3cm]{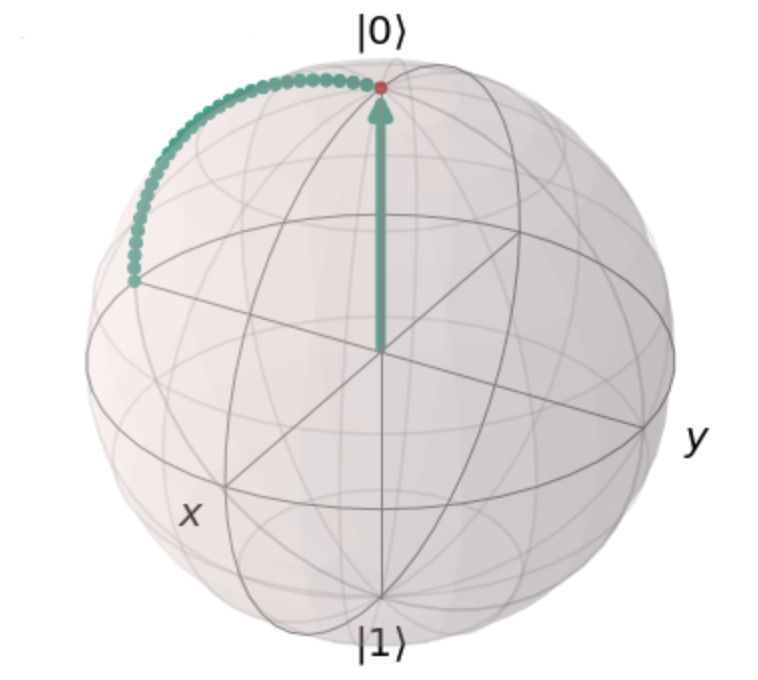}
    \caption{$(R_x\frac{\pi}{2})(CZ_2)(R_x-\frac{\pi}{2}) = 0$}
    \end{subfigure}
    \caption{The Bloch sphere visualizations for the quantum circuit $(R_x\frac{\pi}{2})(CZ_2)(R_x-\frac{\pi}{2})$ applied to a target qubit of state $|0\rangle$, when: (a) $x_2= 1$, and (b) $x_2 = 0$.}
    \label{fig:RZRsphere0}
\end{figure}

Local transformations are now applied to reduce the cost of this quantum circuit. Note that (i) the cost of a quantum circuit in this paper is defined as the total number of quantum gates ($RX$ and $CZ$), and (ii) $(R_x\frac{\pi}{2})(CZ_i)(R_x-\frac{\pi}{2}) = x_i$ when applied to a target qubit of state $|0\rangle$, as illustrated in Figure \ref{fig:RZRsphere0}.

\begin{figure}[!ht]
    \centering
    \scalebox{0.9}{
    \begin{quantikz}
    & \lstick{$x_1$} & \gate{R_X(\frac{\pi}{2})} & \gate{Z} & \gate{R_X(-\frac{\pi}{2})} & \qw \rstick{$f$} \\
    & \lstick{$x_2$} & \qw & \ctrl{-1} & \qw & \qw 
    \end{quantikz}}
    \caption{The quantum circuit for the two-input XOR function, where $x_1$ and $ x_2$ are the input qubits, and $f$ is the final output qubit of a state $|0\rangle$ (as a non-solution) or $|1\rangle$ (as a solution). Note that no ancilla (auxiliary qubit) is utilized in this lower-cost quantum circuit of only three gates.}
    \label{XORcircuit2}
\end{figure}
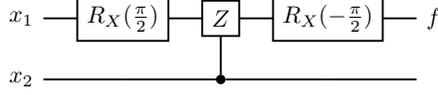

Hence, the first two $(R_x\frac{\pi}{2})(CZ_1)$ gates in Figure \ref{XORcircuit1} can be omitted and replaced with one $(R_x\frac{\pi}{2})$ gate applied on the first qubit ($x_1$). The $x_1$ now plays the role of the target qubit. The last two $(CZ_2)(R_x-\frac{\pi}{2})$ gates (in Figure \ref{XORcircuit1}) are then flipped to the first qubit ($x_1$), such that the target qubit will be totally removed as shown in Figure \ref{XORcircuit2}. By removing the target qubit, we now have a lower-cost quantum circuit of this gate.

\newpage

\noindent \textbf{Example 3:} A Toffoli function is generated with the truth vector of $[0,1,0,1,0,1,1,0]^T$. 

\noindent Since \[2^{-3}W_3\vec{F} = [\frac{1}{2}, 0, -\frac{1}{4}, -\frac{1}{4}, 0, 0, -\frac{1}{4}, \frac{1}{4}]^T\] using the three inputs canonical cascade decomposition from [\ref{0}, \ref{2}],\[a^{w_a}g^{x_3}a^{w_b}g^{x_2 + x_3}a^{w_c}g^{x_3}a^{w_d}g^{x_1+x_2 + x_3}a^{w_e}g^{x_3}a^{w_f}g^{x_2 + x_3}a^{w_g}g^{x_3}a^{w_h}g^{x_1 + x_2+x_3}\text{.}\]
Then the reduced canonical cascade before the quantum conversion is obtained as 
\[f(x) = a^{\frac{1}{2}}g^{x_3}a^{-\frac{1}{4}}g^{x_2}a^{-\frac{1}{4}}g^{x_2}g^{x_1}a^{-\frac{1}{4}}g^{x_2}a^{\frac{1}{4}}\text{.}\]

We now convert every $a^n$ element to a quantum rotational gate around the X-axis and every $g^{x_n}$ element to a quantum rotational gate around the Z-axis of the Bloch sphere using $R_x$ and $CZ$ gates, respectively. After this quantum conversion, the quantum equivalent 2-level canonical cascade becomes:
\[f(x) = (R_x\frac{\pi}{2})(CZ_3)(R_x-\frac{\pi}{4})(CZ_2)(R_x-\frac{\pi}{4})(CZ_2)(CZ_1)(R_x-\frac{\pi}{4})(CZ_2)(R_x\frac{\pi}{4})\text{.}\]

\begin{figure}[!ht]
    \centering
    \includegraphics[width=11cm]{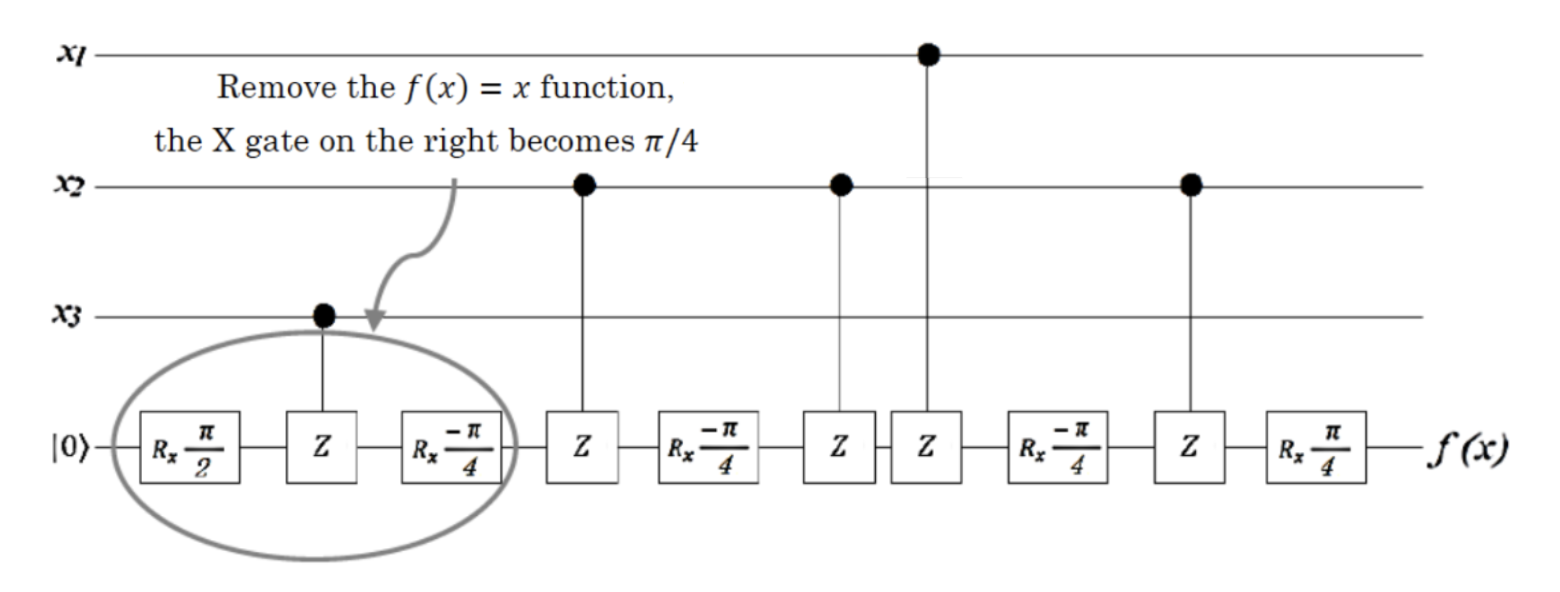}
    \caption{The exact quantum binary (EQB) canonical cascade of a Toffoli gate before applying the reduction, where $x_1$, $x_2$, and $x_3$ are the input qubits, and $f(x)$ is the final output qubit of a state $|0\rangle$ (as a non-solution) or $|1\rangle$ (as a solution).}
    \label{fig:toffoli}
\end{figure}

Next, we apply the local transformations to reduce our canonical cascade, and a significant simplification can be made for the $x_3$ input qubit with the target qubit. Once again since $(R_x\frac{\pi}{2})(CZ_3)(R_x-\frac{\pi}{2}) = x_3$, then the first three gates of the canonical cascade $(R_x\frac{\pi}{2})(CZ_3)(R_x-\frac{\pi}{4})$ can be subtracted out and completely omitted, thereby leaving a remainder gate of $R_x$ of $\frac{\pi}{4}$ rotations. Figure \ref{fig:toffoli} illustrates the quantum circuit of canonical cascade before applying the reduction.

\begin{figure}[!ht]
    \centering
    \includegraphics[width=11cm]{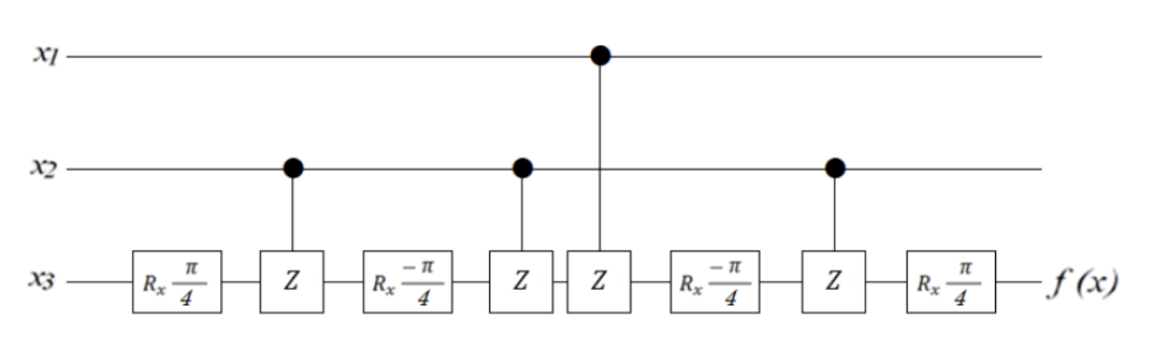}
    \caption{The EQB canonical cascade of a Toffoli gate after applying the reduction, where $x_1$, $x_2$, and $x_3$ are the input qubits, and $f(x)$ is the final output qubit of a state $|0\rangle$ (as a non-solution) or $|1\rangle$ (as a solution). Note that no ancilla qubit is utilized in this lower-cost quantum circuit of only eight gates.}
    \label{fig:toffolired}
\end{figure}

Since $x_3$ has only one control point to the target qubit, $x_3$ can safely play the role of the target qubit. This is a significant reduction, since it removes all ancilla qubits and reduces the cost to only eight gates. Figure \ref{fig:toffolired} depicts the quantum circuit of canonical cascade after applying the reduction.

Notice that, in this paper, all single-qubit gates have only one target qubit, and all double-qubit gates have one control qubit and one target qubit, which gives these gates the advantage of being well-suited to different quantum layouts.

\begin{figure}[!ht]
    \centering
    \includegraphics[width=5.5cm]{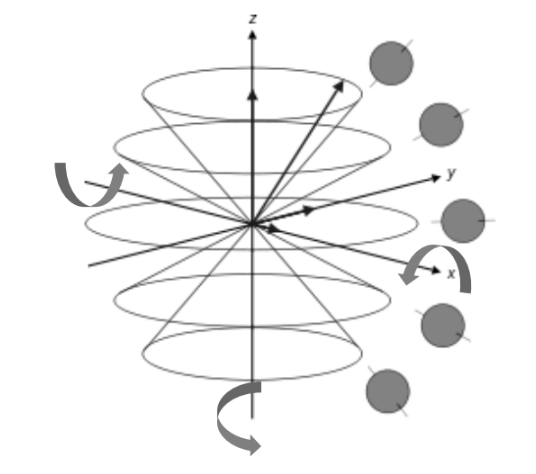}
    \caption{The Bloch sphere visualization for the EQB controlled precession of a qubit during arbitrary gates execution. The CZ gates rotate this qubit around the Z-axis, while the $R_x$ gates change the rotational precession around the X-axis.}
    \label{fig:rotationfig}
\end{figure}

The repetition of CZ gates spins a qubit around the Z-axis of the Bloch sphere for a number of rotations, and this number of rotations is determined by the initial configuration of the input qubits. The rotations of $R_x$ gates cause a qubit to change the spin orientation, which is equivalent to a controlled precession. Finally, a qubit will settle at one of its observable states. Figure \ref{fig:rotationfig} demonstrates how the controlled precession of a qubit may appear in the Bloch sphere during the execution of arbitrary gates, using the group decomposition method with the exact quantum binary (EQB) output.

We emphasize that there is a strong dependency on the size of angular rotations and the number of input variables, via $A_{rot} = 2^{-n}W_n\vec{F}$. This dependency suggests that the number of input variables is limited to the smallest accurate rotations that a quantum hardware can handle. For instance, if a quantum hardware is capable of $0.7^\circ$ rotations, with an accuracy of less than $0.35^\circ$, then the maximum number of inputs that can safely be utilized is approximately $ n = 8$. Depending on the function. this is the worst case where there are no canonical cascade reductions. Recently, quantum technologies, such as IonQ, can handle $RX$ and $RY$ gates with a rotational precision of $\pi \cdot 10^{-3}$ [\ref{9}].



\subsection{Local Transformations}

The local transformations are applied to our proposed methodology of canonical cascades as presented in the following steps.

\begin{enumerate}
    \item Remove all $CZ$ gates at the beginning of a quantum circuit for a canonical cascade, as shown in Figure \ref{fig:CZstart}.

    \begin{figure}[!ht]
        \centering
        \includegraphics[width=6cm]{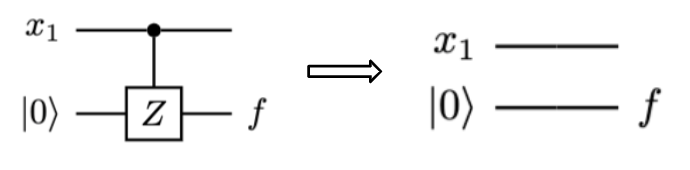}
        \caption{The simplification of $(CZ)$ gate at the beginning of a quantum circuit, where $x_1$ is the input qubit and $f$ is the final output qubit.}
        \label{fig:CZstart}
    \end{figure}

    \item Remove all $CZ$ gates at the end of a quantum circuit for a canonical cascade, as shown in Figure \ref{fig:CZend}.

    \begin{figure}[!ht]
        \centering
        \includegraphics[width=5cm]{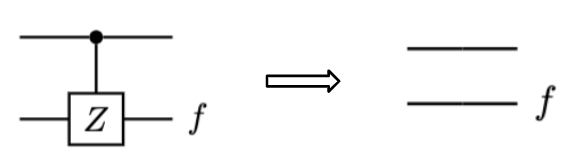}
        \caption{The simplification of $(CZ)$ gate at the end of a quantum circuit, where $f$ is the final output qubit.}
        \label{fig:CZend}
    \end{figure}

    \item When the target qubit is in a state $|1\rangle$, remove any $(R_x\frac{\pi}{2})(CZ_{x_i})(R_x\theta)$ gates in a quantum circuit for a canonical cascade, and replace them with one CNOT gate and one $(R_x\theta + \frac{\pi}{2})$ gate on $x_i$, as shown in Figure \ref{fig:RZRmiddle}, where $i$ is the index of an input qubit.

    \begin{figure}[!ht]
        \centering
        \includegraphics[width=8.5cm]{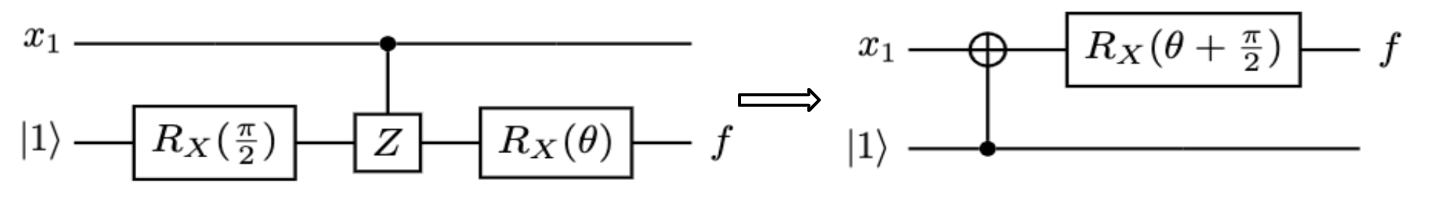}
        \caption{The simplification for a quantum circuit by replacing $(R_x\frac{\pi}{2})(CZ_{x_1})(R_x\theta)$ gates to one CNOT gate and one $(R_x\theta + \frac{\pi}{2})$ gate on the $x_1$ input qubit, where $f$ is the final output qubit.}
        \label{fig:RZRmiddle}
    \end{figure}

     \item When the target qubit is in a state of $|0\rangle$, remove any $(R_x\frac{\pi}{2})(CZ_{x_i})(R_x\theta)$ gates in a quantum circuit for a canonical cascade, and replace them with a $(R_x\theta + \frac{\pi}{2})$ gate on $x_i$, as shown in Figure \ref{fig:RZRmiddle}, where $i$ is the index of an input qubit.

      \begin{figure}[!ht]
        \centering
        \includegraphics[width=8.5cm]{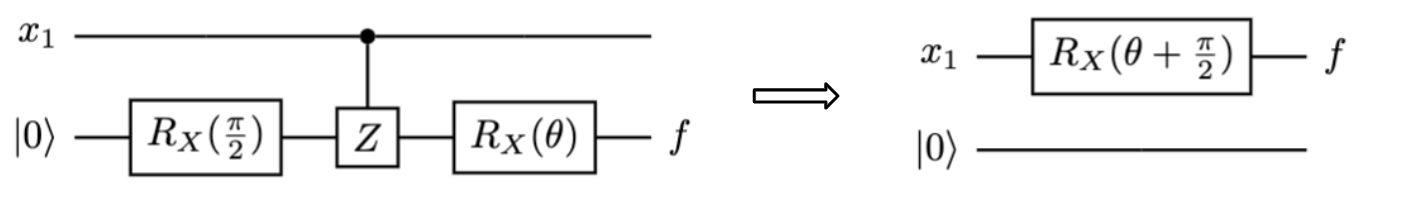}
        \caption{The simplification for a quantum circuit by replacing $(R_x\frac{\pi}{2})(CZ_{x_1})(R_x\theta)$ gates to a$(R_x\theta + \frac{\pi}{2})$ gate on the $x_1$ input qubit, where $f$ is the final output qubit.}
        \label{fig:RZRmiddle}
    \end{figure}

\end{enumerate}

\subsection{Illustrative Examples}

\textbf{Example 4:} Consider the 4-by-4 Peres gate [\ref{17}], i.e., it has four inputs ($x_i$) and four outputs ($f_i$), with its truth table expressed in Table \ref{Peres4table}.

 \begin{table}[!h]
    \centering
        \caption{The truth table of 4-by-4 Peres gate, where $x_i$ is the input, $f_i$ is the output, and $1 \leq i \leq 4$.}

    \begin{tabular}{|c|c|c|c|c|c|c|c|}
    \textbf{$x_1$} & \textbf{$x_2$} & \textbf{$x_3$} & \textbf{$x_4$} & \textbf{$f_1$} & \textbf{$f_2$} & \textbf{$f_3$} & \textbf{$f_4$}\\

    \hline
    0 & 0 & 0 & 0 & 0 & 0 & 0 & 0\\
    0 & 0 & 0 & 1 & 0 & 0 & 0 & 1\\
    0 & 0 & 1 & 0 & 0 & 0 & 1 & 0\\
    0 & 0 & 1 & 1 & 0 & 0 & 1 & 1\\
    0 & 1 & 0 & 0 & 0 & 1 & 0 & 0\\
    0 & 1 & 0 & 1 & 0 & 1 & 0 & 1\\
    0 & 1 & 1 & 0 & 0 & 1 & 1 & 0\\
    0 & 1 & 1 & 1 & 0 & 1 & 1 & 1\\
    1 & 0 & 0 & 0 & 1 & 1 & 0 & 0\\
    1 & 0 & 0 & 1 & 1 & 1 & 0 & 1\\
    1 & 0 & 1 & 0 & 1 & 1 & 1 & 0\\
    1 & 0 & 1 & 1 & 1 & 1 & 1 & 1\\
    1 & 1 & 0 & 0 & 1 & 0 & 1 & 0\\
    1 & 1 & 0 & 1 & 1 & 0 & 1 & 1\\
    1 & 1 & 1 & 0 & 1 & 0 & 0 & 1\\
    1 & 1 & 1 & 1 & 1 & 0 & 0 & 0
    \end{tabular}
    \label{Peres4table}
\end{table}

\noindent Let us first consider $f_1$. The truth vector for this function is $[0,0,0,0,0,0,0,0,1,1,1,1,1,1,1,1]^T$. The reduced canonical cascade for $f_1$ before the quantum conversion is $a^{\frac{1}{2}}g^{x_1}a^{-\frac{1}{2}}$. Hence, after the quantum conversion, its quantum equivalent canonical cascade becomes
\[(R_x\frac{\pi}{2})(CZ_1)(R_x-\frac{\pi}{2})\text{.}\]

\noindent Similarly, the quantum canonical cascade for $f_2$ is \[(R_x\frac{\pi}{2})(CZ_1)(CZ2)(R_x-\frac{\pi}{2})\text{,}\] the quantum canonical cascade for $f_3$ is \[(R_x\frac{\pi}{2})(CZ_3)(R_x-\frac{\pi}{4})(CZ_2)(R_x-\frac{\pi}{4})\text{,}\] and the quantum canonical cascade for $f_4$ is

\begin{center}
$(R_x\frac{\pi}{2})(CZ_4)(R_x\frac{13\pi}{16})(CZ_3)(R_x\frac{15\pi}{16})(CZ_3)(CZ_2)(R_x\frac{13\pi}{16})(CZ_3)(R_x\frac{\pi}{16})(CZ_3)$ \\ $\times (CZ_2)(CZ_1)(R_x\frac{13\pi}{16})(CZ_3)(R_x\frac{\pi}{16})(CZ_3)(CZ_2)(R_x\frac{\pi}{16})(CZ_3)(R_x\frac{13\pi}{16})$.
\end{center}

\begin{figure}[!ht]
    \centering
    \scalebox{0.9}{
    \begin{quantikz}
    &\lstick{$x_1$} & \qw & \ctrl{1} & \qw & \qw \\
    & \lstick{$|0\rangle$} & \gate{R_X(\frac{\pi}{2})} & \gate{Z} & \gate{R_X(-\frac{\pi}{2})} & \qw \rstick{$f_1$}
    \end{quantikz}}
    \caption{The quantum circuit for $f_1$.}
    \label{peres11}
    \end{figure}
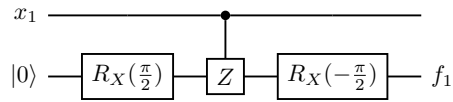

\noindent Therefore, the quantum circuit for $f_1$ is shown in Figure \ref{peres11}.

\begin{figure}[!ht]
    \centering
    \begin{quantikz}
    &\lstick{$x_1$}  & \qw & \qw \rstick{$f_1$}
    \end{quantikz}
    \caption{The simplified quantum circuit for $f_1$.}
    \label{peres12}
    \end{figure}
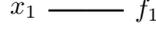

After using the local transformations, this circuit is simplified to its equivalent circuit as shown in Figure \ref{peres12}.

\begin{figure}[!ht]
    \centering
    \scalebox{0.6}{
    \begin{quantikz}
    &\lstick{$x_1$}  & \qw & \qw & \qw & \qw & \qw & \qw & \qw & \qw \qw & \qw & \qw & \ctrl{3} & \qw & \qw & \qw & \qw & \qw & \qw & \qw & \qw & \qw \\
    &\lstick{$x_2$}  & \qw & \qw & \qw & \qw & \ctrl{2} & \qw & \qw & \qw & \qw & \ctrl{2} & \qw & \qw & \qw & \qw & \qw & \ctrl{2} & \qw & \qw & \qw & \qw \\
    &\lstick{$x_3$}  & \qw & \ctrl{1} & \qw & \ctrl{1} & \qw & \qw & \ctrl{1} & \qw & \ctrl{1} & \qw & \qw & \qw & \ctrl{1} & \qw & \ctrl{1} & \qw & \qw & \ctrl{1} & \qw & \qw \\
    &\lstick{$x_4$}  & \gate{R_X(\frac{21\pi}{16})} &  \gate{Z} & \gate{R_X(\frac{15\pi}{16})} & \gate{Z} & \gate{Z} & \gate{R_X(\frac{13\pi}{16})} & \gate{Z} & \gate{R_X(\frac{\pi}{16})} & \gate{Z} & \gate{Z} & \gate{Z} & \gate{R_X(\frac{13\pi}{16})} & \gate{Z} & \gate{R_X(\frac{\pi}{16})} & \gate{Z} & \gate{Z} & \gate{R_X(\frac{\pi}{16})} & \gate{Z} & \gate{R_X(\frac{13\pi}{16})} & \qw \rstick{$f_4$}
    \end{quantikz}}
    \caption{The simplified quantum circuit for $f_4$.}
    \label{peres4}
    \end{figure}
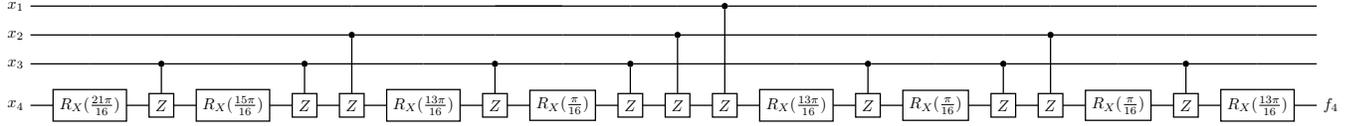

\noindent Similarly, the simplified quantum circuits for $f_2, f_3,$ and $f_4$  after utilizing the local transformations are illustrated in Figure \ref{peres2}, Figure \ref{peres3}, and Figure \ref{peres4}, respectively.

\begin{figure}[!ht]
    \centering
    \begin{quantikz}
    &\lstick{$x_1$}  & \ctrl{1} & \qw \\
    &\lstick{$x_2$}  & \targ{} & \qw \rstick{$f_2$}
    \end{quantikz}
    \caption{The simplified quantum circuit for $f_2$.}
    \label{peres2}
    \end{figure}
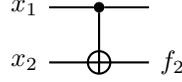

\begin{figure}[!ht]
    \centering
    \begin{quantikz}
    &\lstick{$x_2$}  & \qw & \ctrl{1} & \qw & \qw \\
    &\lstick{$x_3$}  & \gate{R_X(\frac{\pi}{4})} &  \gate{Z} & \gate{R_X(-\frac{\pi}{4})} & \qw \rstick{$f_3$}
    \end{quantikz}
    \caption{The simplified quantum circuit for $f_3$.}
    \label{peres3}
    \end{figure}
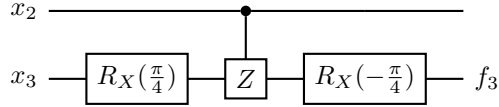

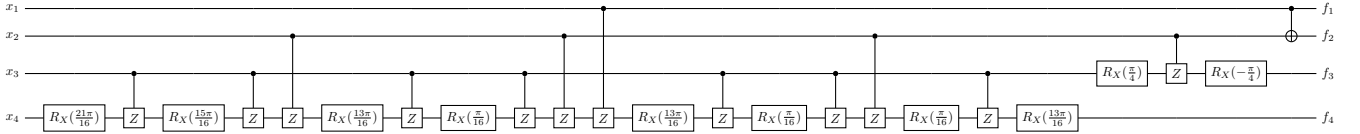
\begin{figure}[!ht]
    \centering
    \scalebox{0.5}{
    \begin{quantikz}
    &\lstick{$x_1$}  & \qw & \qw & \qw & \qw & \qw & \qw & \qw & \qw \qw & \qw & \qw & \ctrl{3} & \qw & \qw & \qw & \qw & \qw & \qw & \qw & \qw & \qw & \qw & \qw & \ctrl{1} & \qw \rstick{$f_1$}\\
    &\lstick{$x_2$}  & \qw & \qw & \qw & \qw & \ctrl{2} & \qw & \qw & \qw & \qw & \ctrl{2} & \qw & \qw & \qw & \qw & \qw & \ctrl{2} & \qw & \qw & \qw & \qw & \ctrl{1} & \qw & \targ{} & \qw \rstick{$f_2$}\\
    &\lstick{$x_3$}  & \qw & \ctrl{1} & \qw & \ctrl{1} & \qw & \qw & \ctrl{1} & \qw & \ctrl{1} & \qw & \qw & \qw & \ctrl{1} & \qw & \ctrl{1} & \qw & \qw & \ctrl{1} & \qw & \gate{R_X(\frac{\pi}{4})} & \gate{Z} & \gate{R_X(-\frac{\pi}{4})} & \qw & \qw \rstick{$f_3$}\\
    &\lstick{$x_4$}  & \gate{R_X(\frac{21\pi}{16})} &  \gate{Z} & \gate{R_X(\frac{15\pi}{16})} & \gate{Z} & \gate{Z} & \gate{R_X(\frac{13\pi}{16})} & \gate{Z} & \gate{R_X(\frac{\pi}{16})} & \gate{Z} & \gate{Z} & \gate{Z} & \gate{R_X(\frac{13\pi}{16})} & \gate{Z} & \gate{R_X(\frac{\pi}{16})} & \gate{Z} & \gate{Z} & \gate{R_X(\frac{\pi}{16})} & \gate{Z} & \gate{R_X(\frac{13\pi}{16})} & \qw & \qw & \qw & \qw & \qw \rstick{$f_4$}
    \end{quantikz}}
    \caption{The final simplified quantum circuit of the 4-by-4 Peres gate, using our approach of reversely combining the four simplified quantum circuits for $f_1$, $f_2$, $f_3$, and $f_4$.}
    \label{peresfinal}
    \end{figure}

By reversely combining the four simplified quantum circuits for $f_1$, $f_2$, $f_3$, and $f_4$, the final simplified quantum circuit of the 4-by-4 Peres gate is constructed, as shown in Figure \ref{peresfinal}. Note that (i) we prefer to use the approach of reversely combining quantum circuits, since such an approach avoids using the mirrors (as the upcomputing parts) of some quantum circuits, (ii) the final reversely combined quantum circuit will always have a lower cost (as a fewer number of quantum gates) than that of the final straightly combined quantum circuit, and (iii) the output functions ($f$) will be the same whether using the reversely or straightly combining approaches.

\noindent \textbf{Example 5:} Consider the controlled-SWAP (as the two-output Fredkin) gate, with its truth table represented in Table \ref{fig:fredkintable}. Note that this gate has three inputs ($x_i$) and two outputs ($\vec{F_j}$).
\newpage
\begin{table}[!ht]
    \begin{center}
        \caption{The truth table of controlled-SWAP (as two-output Fredkin) gate, where $x_i$ is the input, $\vec{F_j}$ is the output, $1 \leq i \leq 3$, and $1 \leq j \leq 2$.}

    \begin{tabular}{|c|c|c|c|c|}
    \textbf{$x_{1}$} & \textbf{$x_{2}$} & \textbf{$x_{3}$} & \textbf{$\vec{F_1}$} & \textbf{$\vec{F_2}$}\\

    \hline
    0 & 0 & 0 & 0 & 0\\
    0 & 0 & 1 & 0 & 1\\
    
    0 & 1 & 0 & 1 & 0\\
    0 & 1 & 1 & 1 & 1\\
    
    1 & 0 & 0 & 0 & 0\\
    1 & 0 & 1 & 1 & 0\\
    
    1 & 1 & 0 & 0 & 1\\
    1 & 1 & 1 & 1 & 1
    \end{tabular}
    \label{fig:fredkintable}
    \end{center}
\end{table}

\begin{figure}[h!t]
    \centering
    \includegraphics[width=4cm]{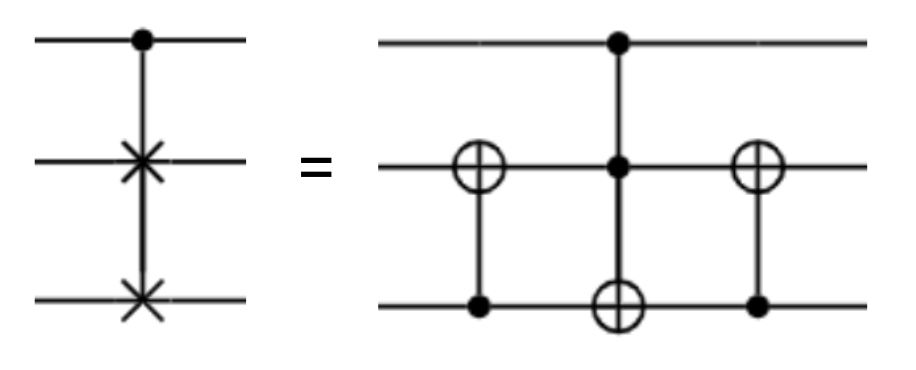}
    \caption{Two representations of Fredkin (controlled-SWAP) gate.}
    \label{Fredkinother}
\end{figure}

\noindent Note that this Fredkin gate is equivalent to one CNOT, one Toffoli, and one CNOT gates, as illustrated in Figure \ref{Fredkinother}.

\begin{figure}[!ht]
    \centering
    \includegraphics[width=11cm]{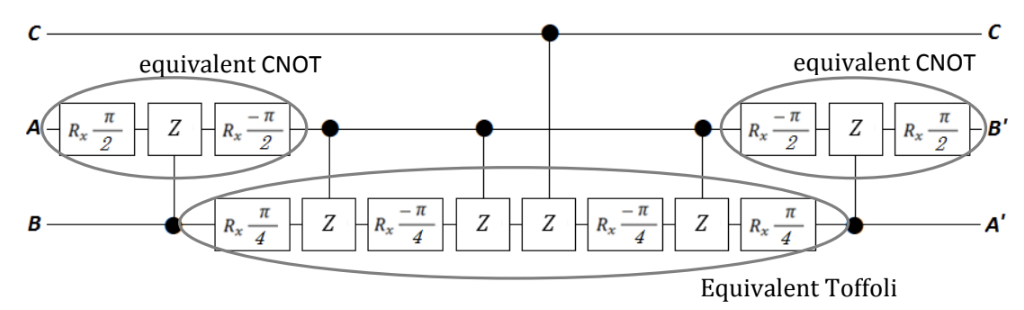}
    \caption{The simplified quantum circuit for the Fredkin gate, which replaces its CNOT and Toffoli gates with their quantum circuits derived in Examples 2 and Example 3 above, respectively.}
    \label{FredkinNoCNOT}
\end{figure}

By replacing the CNOT and Toffoli gates with their quantum circuits derived in Examples 2 and Example 3 above, respectively, we will then obtain the simplified quantum circuit for the Fredkin gate, as shown in Figure \ref{FredkinNoCNOT}.

\noindent \textbf{Example 6:}  To demonstrate the realization of the binary input multi-valued output function using the multiple outputs function, let $f(x_1, x_2, x_3) = x_1+x_2+x_3$, which is a 3-digit adder, and the truth table of this adder is stated in Table \ref{mod3addertable}.

\begin{table}[H]
    \centering
        \caption{The truth table for a 3-digit adder of three binary inputs ($x_i$) and one ternary output $(f(x_i))$, where $1 \leq i \leq 3$.}
    \begin{tabular}{|c|c|c|c|}
    \textbf{$x_3$} & \textbf{$x_2$} & \textbf{$x_1$} & \textbf{$f(x_1, x_2, x_3)$} \\

    \hline
    0 & 0 & 0 & 0\\
    0 & 0 & 1 & 1\\
    0 & 1 & 0 & 1\\
    0 & 1 & 1 & 2\\
    1 & 0 & 0 & 1\\
    1 & 0 & 1 & 2\\
    1 & 1 & 0 & 2\\
    1 & 1 & 1 & 3
    \end{tabular}
    \label{mod3addertable}
\end{table}

\noindent Hence, the truth vector for this 3-digit adder is $\vec{F} = [0, 1, 1, 2, 1, 2, 2, 3]^T$, and the mapping of this adder is shown in Figure \ref{mod3addermap}.

\begin{figure}[!ht]
\begin{center}
\scalebox{.85}{
    \begin{tikzpicture}[karnaugh, American style, kmlabel left/.style={black, left=0pt}, kmlabel top/.style={black, yshift=8pt}]
  \karnaughmaptabvert{3}{}{{$x_3$}{$x_1$}{$x_2$}}%
     {{0}{1}
      {2}{1}
      {1}{2}
      {3}{2}
      }{}
\end{tikzpicture}}
\caption{Map for the function of a 3-digit adder: $f(x_{1}, x_{2}, x_{3}) = x_{1} + x_{2} + x_{3}$.}
\label{mod3addermap}
\end{center}
\end{figure}

\noindent To convert a ternary value to a set of binary values, we perform the following binary map: $0 = 00_2$, $1  = 01_2$, $2 = 10_2$, and $3 = 11_2$, then we replace the ternary values in Figure \ref{mod3addermap} to this set of binary values, as shown in Figure \ref{mod3adderbinmap}.

\begin{figure}[!ht]
\begin{center}
\scalebox{0.9}{
    \begin{tikzpicture}[karnaugh, American style, kmlabel left/.style={black, left=0pt}, kmlabel top/.style={black, yshift=8pt}]
  \karnaughmaptabvert{3}{}{{$x_3$}{$x_1$}{$x_2$}}%
     {{00}{01}
      {10}{01}
      {01}{10}
      {11}{10}
      }{}
\end{tikzpicture}}
\caption{Binary map for the function of a 3-digit adder: $f(x_{1}, x_{2}, x_{3}) = x_{1} + x_{2} + x_{3}$.}
\label{mod3adderbinmap}
\end{center}
\end{figure}

\noindent Now, we can split this binary map into two standard binary maps, by considering the first and second digits separately, as shown in Figure \ref{2mod3adderbinmap}.

\begin{figure}[!ht]
\centering
\begin{subfigure}{.5\textwidth}
  \centering
  \scalebox{0.9}{
   \begin{tikzpicture}[karnaugh, American style, kmlabel left/.style={black, left=0pt}, kmlabel top/.style={black, yshift=8pt}]
  \karnaughmaptabvert{3}{}{{$x_3$}{$x_1$}{$x_2$}}%
     {{0}{0}
      {1}{0}
      {0}{1}
      {1}{1}
      }{}
\end{tikzpicture}}
\caption{Sub-map for the first digit}
\label{mod3adderbinmapa}
\end{subfigure}%
\begin{subfigure}{.5\textwidth}
  \centering
  \scalebox{0.9}{
   \begin{tikzpicture}[karnaugh, American style, kmlabel left/.style={black, left=0pt}, kmlabel top/.style={black, yshift=8pt}]
  \karnaughmaptabvert{3}{}{{$x_3$}{$x_1$}{$x_2$}}%
     {{0}{1}
      {0}{1}
      {1}{0}
      {1}{0}
      }{}
\end{tikzpicture}}
\caption{Sub-map for the second digit}
\label{mod3adderbinmapb}
\end{subfigure}
\caption{Two sub-maps of the binary map for the function of a 3-digit adder: $f(x_{1}, x_{2}, x_{3}) = x_{1} + x_{2} + x_{3}$.}
\label{2mod3adderbinmap}
\end{figure}
\newpage
\noindent Let us consider these two sub-maps as two different functions; namely, functions $f_1$ and $f_2$. Since $2^{-3}W_3\vec{f_1} = [\frac{1}{2}, -\frac{1}{4}, -\frac{1}{4}, 0, -\frac{1}{4}, 0, 0, \frac{1}{4}]^T$, the reduced canonical cascade for $f_1$ before the quantum conversion is obtained as
\[a^{\frac{1}{2}}g^{x_3}a^{-\frac{1}{4}}g^{x_2 + x_3}a^{-\frac{1}{4}}g^{x_1+x_2}a^{-\frac{1}{4}}g^{x_2+x_3}a^{\frac{1}{4}}\text{.}\]

\noindent Thus, after the quantum conversion, the simplified quantum circuit of $f_1$ becomes
\[(R_x\frac{\pi}{2})(CZ_3)(R_x-\frac{\pi}{4})(CZ_3)(CZ_2)(R_x-\frac{\pi}{4})(CZ_2)(CZ_1)(R_x-\frac{\pi}{4})(CZ_3)(CZ_2)(R_x\frac{\pi}{4})\text{.}\]

\noindent Since $2^{-3}W_3\vec{f_2} = [\frac{1}{2},0,0, 0,0, 0, 0, -\frac{1}{2}]^T$, the reduced canonical cascade for $f_2$ before the quantum conversion is obtained as
\[a^{\frac{1}{2}}g^{x_1+x_2 + x_3}a^{-\frac{1}{2}}\text{.}\]

\noindent Thus, after the quantum conversion, the simplified quantum circuit of $f_2$ becomes
\[(R_x\frac{\pi}{2})(CZ_3)(CZ_2)(CZ_1)(R_x-\frac{\pi}{2})\text{.}\]

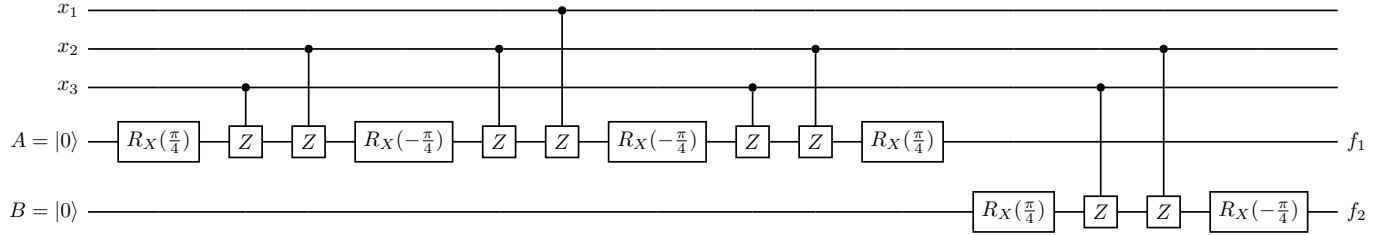
\begin{figure}[!ht]
    \centering
    \scalebox{0.8}{
    \begin{quantikz}
    & \lstick{$x_1$} & \qw & \qw & \qw & \qw & \qw & \ctrl{3} & \qw & \qw & \qw & \qw & \qw & \qw & \qw & \qw & \qw \\
    & \lstick{$x_2$} & \qw & \qw & \ctrl{2} & \qw & \ctrl{2} & \qw & \qw & \qw & \ctrl{2} & \qw & \qw & \qw & \ctrl{3} & \qw & \qw \\
    & \lstick{$x_3$} & \qw & \ctrl{1} & \qw & \qw & \qw & \qw & \qw & \ctrl{1} & \qw & \qw & \qw & \ctrl{2} & \qw & \qw & \qw \\
    & \lstick{$A = |0\rangle$} & \gate{R_X(\frac{\pi}{4})} & \gate{Z} & \gate{Z} & \gate{R_X(-\frac{\pi}{4})} & \gate{Z} & \gate{Z} & \gate{R_X(-\frac{\pi}{4})} & \gate{Z} & \gate{Z} & \gate{R_X(\frac{\pi}{4})} & \qw & \qw & \qw & \qw & \qw  \rstick{$f_1$} \\
    & \lstick{$B = |0\rangle$} & \qw & \qw & \qw & \qw & \qw & \qw & \qw & \qw & \qw & \qw & \gate{R_X(\frac{\pi}{4})} & \gate{Z} & \gate{Z} & \gate{R_X(-\frac{\pi}{4})} & \qw \rstick{$f_2$}
    \end{quantikz}}
    \caption{The final simplified quantum circuit for the 3-digit adder, where $x_1$, $ x_2$, and $x_3$ are the input qubits, A and B are the target qubits initially set to the state of $|0\rangle$, and $f_1$ and $f_2$ are the final output qubits of a state $|0\rangle$ or $|1\rangle$.}
    \label{3adder}
\end{figure}

\noindent After applying the local transformations, the final quantum circuit for the 3-digit adder is constructed, as shown in Figure \ref{3adder}.

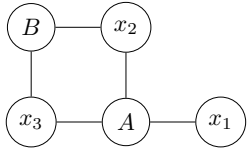
\begin{figure}[!ht]
    \centering
    \scalebox{0.9}{
    \begin{tikzpicture}
\node[circle,draw, minimum size=0.7cm] (A) at  (0,0) {$A$};
\node[circle,draw, minimum size=0.7cm] (1) at  (1.4,0)  {$x_1$};
\node[circle,draw, minimum size=0.7cm] (3) at  (-1.4,0) {$x_3$};
\node[circle,draw, minimum size=0.7cm] (2) at  (0,1.4) {$x_2$};
\node[circle,draw, minimum size=0.7cm] (B) at  (-1.4,1.4) {$B$};

\draw (A) -- (3);
\draw (A) -- (1);
\draw (A) -- (2);
\draw (B) -- (2);
\draw (B) -- (3);

\end{tikzpicture}}

\caption{A quantum layout (as a square lattice) for a better synthesizing of a 3-digit adder circuit, where $x_1$, $ x_2$, and $x_3$ are the input qubits, and A and B are the target qubits of this circuit.}
\label{3AdderLayout}
\end{figure}

\noindent Note that the final simplified quantum circuit of this 3-digit adder example can be better synthesized for a quantum layout, since the input and target qubits of this circuit can be re-arranged in a square lattice, as shown in Figure \ref{3AdderLayout}.

\subsection{EQB Synthesis Tool}

\indent Our proposed exact quantum binary (EQB) synthesis tool is a simple program that designs simple quantum circuits and verifies the results. This EQB synthesis tool generates quantum canonical cascades exactly as outlined in Section 3.1 above and provides a solution for classical reversible cascades, as presented in [\ref{0}, \ref{2}].

Recently, our EQB synthesis tool was limited to 10 input variables ($n = 10$). Since this tool generates a string of $3*2^{n}-2$ elements ($a$ and $g$) to represent the group decomposition expression, the processing time for calculating a solution significantly increases as $n$ increases. To employ the EQB synthesis tool, a user needs to provide the number of input variables ($n$), the number of logic levels ($p$), and the completely specified truth table for a function ($truth~vector$). The program produces circuits using the EQB Synthesis method as outlined in section 3.1 or the multivalued group decomposition (MGD) method as illustrated in [\ref{0}, \ref{2}], depending on what the user inputs. If EQB synthesis is desired, the user must enter the value of $p$ in the tool. The program also provides the user with a visual representation of the quantum rotations that are being executed on the target qubit, as the cascade is either executed one gate at a time or entirely at once

The EQB synthesis tool starts by (i) generating the $n$-order Walsh matrix ($W_n$) of size $2^{n} \times 2^n$, where $n \geq 1$, (ii) multiplying $W_n$ by the truth vector of a function, and (iii) deleting all string elements containing zero coefficients, along with consecutive adjacent elements ($g^xa^0g^x = g^xIg^x = I$), where $a^0 = I$. The EQB synthesis tool then generates the inverse Walsh factor ($W_n^{-1}$) based on the fraction $2^{-n}$, with a modulo of $2p$. If EQB is desired, then no modulo is applied. The $W_n^{-1}$ is multiplied by the remaining string elements ($a$ and $g$), to generate the Walsh spectrum ($\vec{w}$), as $\vec{w} = 2^{-n}W_n\vec{F}$. If EQB is requested, our tool checks the truth vector for the symmetry. If symmetry exists, the target qubit becomes an input qubit for the least significant input variable. Such a symmetry allows for the removal of at least one $Z$ gate, one rotational gate ($RX$ and $RY$), and a target ancilla qubit. It should be pointed out that searching for symmetries and responding to certain symmetries almost always seems to result in a reduction of canonical cascades. Currently, the EQB synthesis tool does not perform an exhaustive search for symmetries, symmetries resulting from any $don’t~care$ states, or any other exclusive-or sum of products (ESOP) minimization [\ref{18}]. It is very possible that additional reductions will be discovered if such searches are exhaustively performed on functions with a large number of input variables, but such analyses will be reserved for future works.

\section{Results}

Table \ref{fig:benchmark} summarizes a set of benchmark functions (quantum circuits), which are generated using our proposed EQB synthesis tool, against the quantum costs reported by RevLib [\ref{19}] and Maslov [\ref{20}]. Each benchmark quantum circuit was chosen because its outputs are limited to one or two equivalent 3 input XOR equivalent CNOT equivalent CNOT Equivalent Toffoli. Note that our EQB synthesis tool is not currently optimized for multi-output functions. Table \ref{fig:benchmark} is organized as follows.

\begin{itemize}
    \item EQB Gates Count – Records the primitive gates count reported by the EQB synthesis tool, including all $CZ$ and $RX$ gates.
    \item EQB Quantum Cost – Records the number of controlled gates reported by the EQB synthesis tool, after absorption of all $RX$ gates into neighboring $CZ$ gates (Maslov quantum cost).
    \item  Correction Quantum Cost (CQC) – Records the number of gates needed to detect a phase shift through the control lines, and the number of correction gates needed to correct the phase at each output (on average the cost is $n(2 + m)$, where $n$ is the number of inputs and $m$ is the number of outputs).
    \item  Ancilla Qubits – Records whether an ancilla qubit is used as the final output qubit.
    \item Best Reported – Records the lowest quantum cost reported by the aforementioned references [\ref{19}, \ref{20}].
    \item Worst Reported – Records the highest quantum cost reported by the aforementioned references [\ref{19}, \ref{20}].
    \item  Percent Potential Improvement (PPI) – Records the percentage of reduction using Maslov cost. If there is no reduction, then the "no change" is reported. If the EQB synthesis tool increases the cost, then the ``worse" is reported.
    
\end{itemize}

\begin{table}[!h]
    \centering
        \caption{EQB benchmark comparisons for different functions (quantum circuits) against the quantum costs reported by RevLib [\ref{19}] and Maslov [\ref{20}].}
    \scalebox{.7}{
    \begin{tabular}{|c|c|c|c|c|c|c|c|}
    \text{\CellWithForceBreak{Benchmark Functions}}& \text{EQB Gates  Count}& \text{\CellWithForceBreak{EQB Quantum Cost \\ (Maslov cost)}} & \text{\CellWithForceBreak{CQC}} & \text{Ancilla Qubits}& \text{\CellWithForceBreak{Best Reported \\ from [19], [20]}} & \text{\CellWithForceBreak{Worst Reported \\ from [19], [20]}} & \text{\CellWithForceBreak{PPI}} \\

    \hline
    Fredkin (controlled-SWAP)& 14 & 6 & 4 & 0 & 7 & 15 & 14\\
    Feynman (CX)& 3 & 1 & 3 & 0 & 1 & n/a & no change\\
    3-bit Toffoli& 8 & 4 & 6 & 0 & 5 & n/a & 20\\
    4-bit Toffoli& 19 & 11 & 9 & 0 & 13 & n/a & 15\\
    5-bit Toffoli& 42 & 26 & 12 & 0 & 26 & 29 & 0\\
    6-bit Toffoli& 89 & 57 & 15 & 0 & 38 & 61 & worse\\
    \CellWithForceBreak{4 greater than 10}& 42 & 26 & 9 & 1 & 34 & 53 & 14\\
    \CellWithForceBreak{4 greater than 11}& 8 & 4 & 9 & 1 & 7 & 16 & 43\\
    \CellWithForceBreak{4 greater than 12}& 42 & 26 & 9 & 1 & 41 & 58 & 37\\
    \CellWithForceBreak{4 greater than 13}& 19 & 11 & 9 & 1 & 15 & 34 & 27\\
    \CellWithForceBreak{4 greater than 4}& 42 & 26 & 9 & 1 & 54 & 89 & 52\\
    \CellWithForceBreak{4 greater than 5}& 19 & 11 & 9 & 1 & 21 & 29 & 48
    \end{tabular}}
    \label{fig:benchmark}
\end{table}

\section{Conclusion}

We introduce a new group decomposition methodology based on the Walsh transform to synthesize the canonical cascades into quantum circuits using the quantum gates $(CNOT$, $RX$, $RY$, and $CZ)$, propose a new class of local transformations to reduce the total number of quantum gates for a quantum circuit, and develop a new exact quantum binary (EQB) tool that implements our group decomposition methodology and local transformations to synthesis final reduced quantum circuits. Our proposed approaches for synthesizing reduced quantum circuits have two advantages. First, such circuits can manipulate multi-valued logic, e.g., they can have 2-valued (binary) inputs and 3-valued (ternary) outputs. Second, such circuits have good characteristics for mapping them into different quantum layouts (especially the square lattices), since all final synthesized single-qubit gates have one output qubit, and all final synthesized double-qubit gates have one control qubit and one output qubit. Our future work includes generalizing our approaches (group decomposition methodology, local transformations, and EQB synthesis tool) for ternary quantum logic computing.

\section*{References}

\begin{enumerate}

\item Agarwal, I., Saraivanov, M., \& Perkowski, M. (2024). Synthesis of Binary-Input Multi-Valued Output Optical Cascades for Reversible and Quantum Technologies. arXiv preprint arXiv:2410.18367.
 \label{0}

\item Sasao, T. (2003, May). Cascade realizations of two-valued input multiple-valued output functions using decomposition of group functions. In 33rd International Symposium on Multiple-Valued Logic, 2003. Proceedings. https://doi.org/10.1109/ismvl.2003.1201396 \label{1}

\item Saraivanov, M., \& Perkowski, M. (2018). Multi-valued quantum cascade realization with group decomposition. In 2018 IEEE 48th International Symposium on Multiple-Valued Logic (ISMVL). https://doi.org/10.1109/ismvl.2018.00050 \label{2}

\item Saraivanov, M. S. (2013). Quantum circuit synthesis using group decomposition and Hilbert spaces (Master's thesis, Portland State University, Department of Electrical and Computer Engineering). https://doi.org/10.15760/etd.1108 \label{3}

\item Elspas, B., \& Stone, H. S. (1967). Decomposition of group functions and the synthesis of multirail cascades. In 8th Annual Symposium on Switching and Automata Theory (SWAT 1967). https://doi.org/10 .1109/focs.1967.10 \label{4}

\item Yang, G., Song, X., Hung, W. N., Xie, F., \& Perkowski, M. A. (2006, May). Group theory based synthesis of binary reversible circuits. In International Conference on Theory and Applications of Models of Computation (pp. 365-374). Berlin, Heidelberg: Springer Berlin Heidelberg. \label{5}

\item Yang, G., Song, X., Hung, W. N., \& Perkowski, M. A. (2005, January). Fast synthesis of exact minimal reversible circuits using group theory. In Proceedings of the 2005 Asia and South Pacific Design Automation Conference (pp. 1002-1005). \label{6}

\item Nakajima, Y., Kawano, Y., \& Sekigawa, H. (2005). A new algorithm for producing quantum circuits using KAK decompositions. arXiv preprint quant-ph/0509196. \label{7}

\item Al-Bayaty, A., \& Perkowski, M. (2023). GALA-n: Generic architecture of layout-aware n-bit quantum operators for cost-effective realization on IBM quantum computers. arXiv:2311.06760. \label{8}

\item IonQ. (2024, January 18). Best practices for using IonQ hardware. IonQ. https://ionq.com/docs/best-practices-for-using-ionq-hardware. Accessed 22 May 2024. \label{9}

\item Lukac, M., Nursultan, S., Krylov, G., \& Keszöcze, O. (2020, August). Geometric refactoring of quantum and reversible circuits: Quantum layout. In 2020 23rd Euromicro Conference on Digital System Design (DSD) (pp. 428-435). IEEE. \label{10}

\item Tan, B., \& Cong, J. (2020, November). Optimal layout synthesis for quantum computing. In Proceedings of the 39th International Conference on Computer-Aided Design (pp. 1-9). \label{11}

\item Whitney, M., Isailovic, N., Patel, Y., \& Kubiatowicz, J. (2007, May). Automated generation of layout and control for quantum circuits. In Proceedings of the 4th International Conference on Computing Frontiers (pp. 83-94). \label{12}

\item Perkowski, M., Lukac, M., Shah, D., \& Kameyama, M. (2011). Synthesis of quantum circuits in linear nearest neighbor model using positive Davio lattices. Facta Universitatis, Series: Electronics and Energetics. \label{13}

\item Grigorchuk, R., \& Yang, R. (2017). Joint spectrum and the infinite dihedral group. Proceedings of the Steklov Institute of Mathematics, 297 (pp. 145-178). \label{14}

\item Yoeli, M., \& Turner, J. (1967). Decompositions of group functions with applications to two-rail cascades. Information and Control, 10(6) (pp. 565-571). https://doi.org/10.1016/s0019-9958(67)91024-8 \label{15}

\item Kim, H., Choi, I. S., \& Hwang, S. Y. (1997). Design of heuristic algorithms based on Shannon expansion for low-power logic circuit synthesis. IEE Proceedings - Circuits, Devices and Systems, 144(6) (pp. 355-360). \label{16}

\item Alasow, A., Jin, P., \& Perkowski, M. (2022). Quantum algorithm for variant maximum satisfiability. Entropy, 24(11) (p. 1615). \label{17}

\item Kozlowski, T., Dagless, E. L., \& Saul, J. M. (1995, October). An enhanced algorithm for the minimization of exclusive-or sum-of-products for incompletely specified functions. In Proceedings of ICCD'95 International Conference on Computer Design. VLSI in Computers and Processors (pp. 244-249). IEEE. \label{18}

\item Wille, R., Große, D., Teuber, L., Dueck, G. W., \& Drechsler, R. (2008). RevLib: An online resource for reversible functions and reversible circuits. In International Symposium on Multi-Valued Logic (pp. 220-225). \label{19}

\item Maslov, D. (2009, September 2). Reversible logic synthesis benchmarks page. [Online]. Available: https://reversiblebenchmarks.github.io/. [Accessed 20 February 2024]. \label{20}

\item Ghosh, Swaroop, Swarup Bhunia, and Kaushik Roy.``Shannon expansion based supply-gated logic for
improved power and testability.” 14th Asian Test Symposium (ATS’05). IEEE, 2005 \label{21}
\end{enumerate}

\newpage
\section*{A Walsh Spectrum}

\setcounter{figure}{0}  
\renewcommand{\thefigure}{A\arabic{figure}}

Let $G$ be a group and $B = \{0,1\}$. Then, $F: B^n \rightarrow G$ is a group function. By Shannon's Expansion [\cite{21}], a group function $F(x):B^n \rightarrow D_3$ decomposes as follows: 
\begin{equation}F(\hat{X},x_n) = F_a(\hat{X})g^{x_n}F_b(\hat{X})g^{x_n}, \label{eq1} \tag{4} \end{equation} 
where $x_n$ is a two-valued input variable and   $F_a(\hat{X})$ and $F_b(\hat{X})$ denote group functions that do not depend on $x_n$ with $\hat{X}=(x_1, x_2, \ldots x_{m-1}) \in X^{m-1}$. A proof can be found in [10]. This decomposition is similar to Shannon’s expansion for a classical binary logic function and is essential in the design of canonical cascades. Figure A1 shows the canonical cascade function of equation 4:

\begin{figure}[!ht]
    \centering
    \includegraphics[width=7cm]{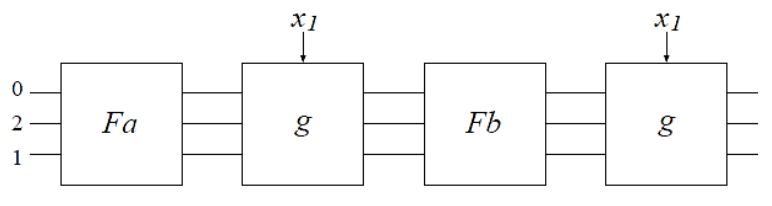}
    \caption{One Variable Canonical Cascade.}
\end{figure}

The decomposition indicates that element $g$ needs to be modified so that it can be controlled with a binary input variable. When $x = 1$, $g^x$ swaps the bottom two lines, and when $x=0$, $g^x$ performs the identity permutation. Figure A2 shows a representation of $g^x$. 

\begin{figure}[!ht]
    \centering
    \includegraphics[width=5cm]{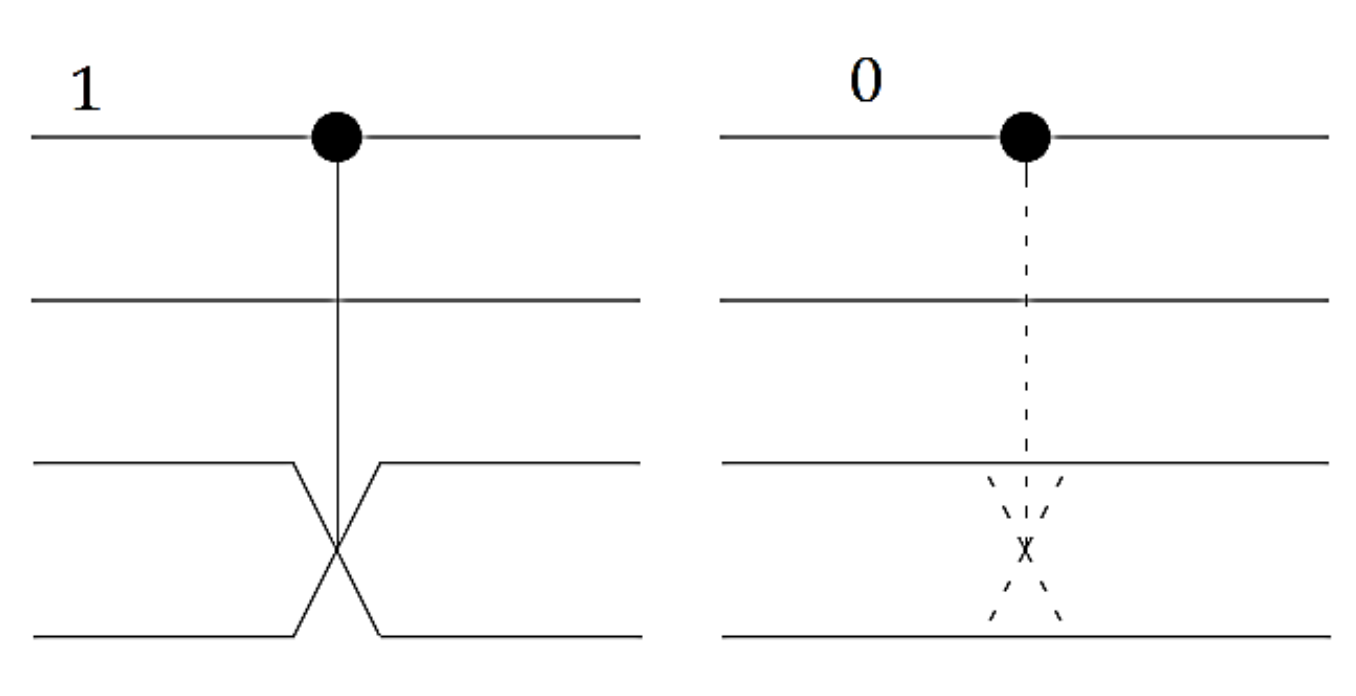}
    \caption{Element $g$ modified to $g^{x_n}$ so it can be controlled.}
\end{figure}

We can let $F_a(\hat{X}) = a^{f_a(\hat{X})}$ and $F_b(\hat{X}) = a^{f_b(\hat{X})}$ where $a$ is the shift element from our group.

From equation 4 we get \[F(\hat{X},x_n) = F_a(\hat{X})g^{x_n}F_b(\hat{X})g^{x_n} = a^{f_a(\hat{X})}g^{x_n}a^{f_b(\hat{X})}g^{x_n}. \label{eq:2} \tag{5}\]

Therefore, \[F(\hat{X},0) = a^{f_a(\hat{X})}g^{0}a^{f_b(\hat{X})}g^{0},\] \[F(\hat{X},1) = a^{f_a(\hat{X})}g^{1}a^{f_b(\hat{X})}g^{1}.\]

Since $g^0 = I$ and $gag = a^{-1}$, we get 
\[F(\hat{X},0) = a^{f_a(\hat{X}) + f_b(\hat{X})},\] \[F(\hat{X},1) = a^{f_a(\hat{X}) - f_b(\hat{X})}.\]

For convenience, we can express the above equations in terms of their exponents only:

\[f(\hat{X},0) = {f_a(\hat{X}) + f_b(\hat{X})},\] \[f(\hat{X},1) = {f_a(\hat{X}) - f_b(\hat{X})}.\]

Converting the above expressions into matrix form we get, 
\[\begin{bmatrix}
f(\hat{X},0) \\
f(\hat{X},1) 
\end{bmatrix} = \begin{bmatrix}
+1 & +1 \\
+1 & -1
\end{bmatrix} \begin{bmatrix}
f_a(\hat{X}) \\
f_b(\hat{X}) 
\end{bmatrix}
\]

\[\Rightarrow \vec{F} = W_1 \vec{w}, \label{eq:3} \tag{6}\]
where $\vec{F}$ is the truth vector of the function and $W_1$ is the first Walsh Matrix. Walsh transform as well as matrix representation of this transform are presented in [8, 13, 21, 22, 57]. Note that $f_a(\hat{X})$ and $f_b(\hat{X})$ represent the exponents of $a$ and literally describe the canonical form of the circuit cascade. If we let $\begin{bmatrix}
f_a(\hat{X}) \\
f_b(\hat{X}) 
\end{bmatrix} = \begin{bmatrix}
w_a \\
w_b 
\end{bmatrix}$, then $w_a$ and  $w_b$ are the exponents of $a$ in our cascade. Hence, the canonical cascade can be found by solving equation 5 for $\vec{w}$ and $\vec{w}$ is the Walsh Spectrum of $\vec{F}$. Multiplying both sides of equation 6 by $W_1^{-1}$, the inverse of the first-order Walsh Matrix, we get, \[\vec{w} = (W_1^{-1}\vec{F}). \label{eq:7} \tag{7}\]

\section*{B Multi-Input Group Decomposition Expression}

\setcounter{figure}{0}  
\renewcommand{\thefigure}{B\arabic{figure}}

Expression 1 can be extended to two variables:
\[F(x_1,x_2) = F_a(x_2)g^{x_1}F_b(x_2)g^{x_1},\]
but in this case $F_a$ and $F_b$ are functions of one variable. Figure B1 shows the canonical cascade:

\begin{figure}[!ht]
    \centering
    \scalebox{0.8}{
    \includegraphics[width=9cm]{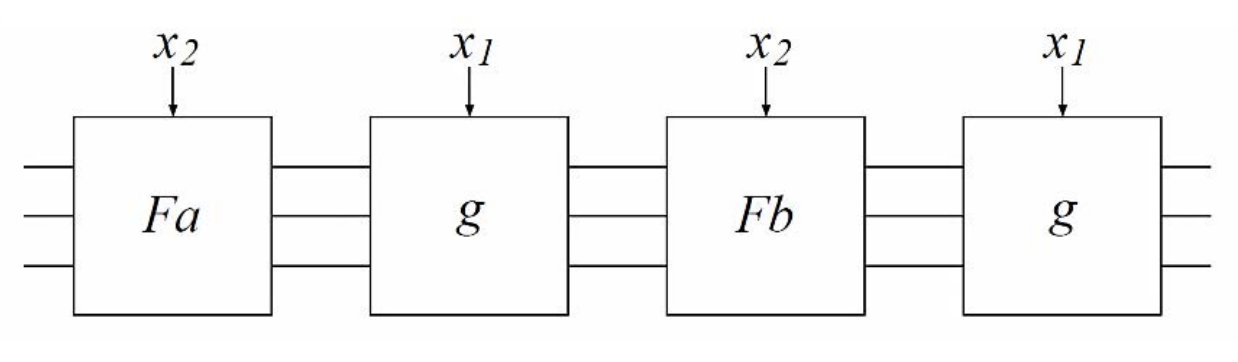}}
    \caption{Intermediate Cascade for two variable functions.}
\end{figure}

From Appendix A we found that a one variable function $F$ can be decomposed as $a^{w_a}g^{x}a^{w_b}g^x$. Since $F_a$ and $F_b$ are functions of one variable, we can replace them with $a^{w_a}g^{x_2}a^{w_b}g^{x_2}$ and $a^{w_c}g^{x_2}a^{w_d}g^{x_2}$ respectively. The canonical form for all functions with two input variables then becomes:

\[a^{f(x_1,x_2)} = ((a^{w_a}g^{x_2}a^{w_b}g^{x_2})g^{x_1})((a^{w_c}g^{x_2}a^{w_d}g^{x_2})g^{x_1}). \label{eq:8} \tag{8}\]

The cascade has ten cells and is shown in Figure B2.

\begin{figure}[H]
    \centering
    \includegraphics[width=10cm]{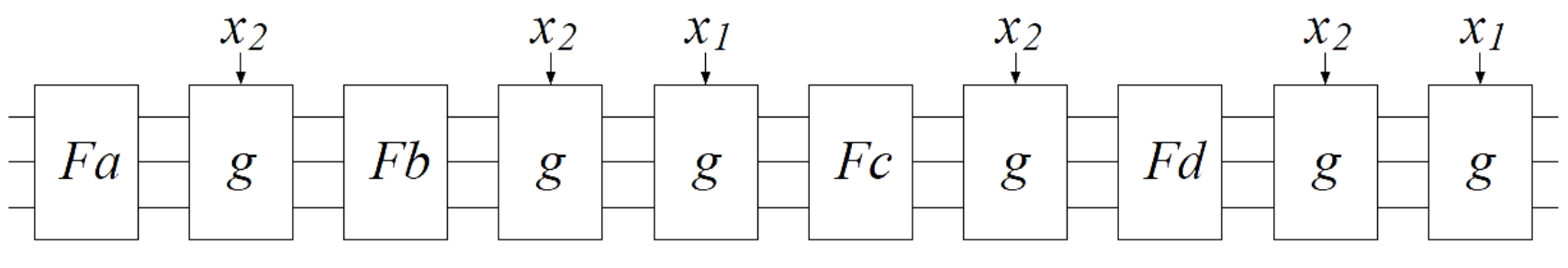}
    \caption{ Cascade for two variable functions after decomposition.}
\end{figure}

Next, we will expand the Walsh matrix to two variables by assigning 0 and 1 to $x_1$ and $x_2$ for all four possible combinations and apply vectors to the exponents as was done in Appendix A for the single variable case. After
applying the property of $gag = a^{-1}$ from Section 2.2, the following four equations are derived:

\begin{center} $a^{f(0,0)} = ((a^{w_a}g^{0}a^{w_b}g^{0})g^{0})((a^{0}g^{x_2}a^{0}g^{x_2})g^{0}) = a^{w_a+w_b+w_c+w_d}$ \\ $\Rightarrow f(0,0) = w_a+w_b+w_c+w_d$, \end{center}
\begin{center} $a^{f(0,1)} = ((a^{w_a}g^{1}a^{w_b}g^{1})g^{0})((a^{w_c}g^{1}a^{w_d}g^{1})g^{0}) = (a^{w_a}a^{-w_b})(a^{w_c}a^{-w_d}) = a^{w_a-w_b+w_c-w_d}$ \\ $\Rightarrow f(0,1) = w_a-w_b+w_c-w_d$, \end{center}
\begin{center} $a^{f(1,0)} = ((a^{w_a}g^{0}a^{w_b}g^{0})g^{1})((a^{w_c}g^{0}a^{w_d}g^{0})g^{1})  = (a^{w_a+w_b})g^{1}(a^{w_c+w_d})g^{1} = a^{w_a+w_b}a^{-(w_c+w_d)} = a^{w_a+w_b-w_c-w_d}$ \\ $\Rightarrow f(1,0) = w_a+w_b-w_c-w_d$, \end{center}
\begin{center} $a^{f(1,1)} = ((a^{w_a}g^{1}a^{w_b}g^{1})g^{1})((a^{w_c}g^{1}a^{w_d}g^{1})g^{1}) = (a^{w_a}a^{-w_b})g^{1}(a^{w_c}a^{-w_d})g^{1}  = a^{w_a-w_b-w_c+w_d}$ \\ $\Rightarrow f(1,1) = w_a-w_b-w_c+w_d$. \end{center}

Putting the above equations in matrix form, we obtain 

\[\begin{bmatrix}
    f(0,0) \\ f(0,1) \\ f(1,0) \\ f(1,1)
\end{bmatrix} = \begin{bmatrix}
    1 & 1 & 1 & 1 \\
    1 & -1 & 1 & -1 \\
    1 & 1 & -1 & -1 \\
    1 & -1 & -1 & 1 
\end{bmatrix}\begin{bmatrix}
    w_a \\ w_b \\ w_c \\ w_d
\end{bmatrix}.\]

Thus, just as in the one variable case, the elements of the $w$ vector in the canonical cascade can be found using the equation 
\[\vec{w} = \begin{bmatrix}
    1 & 1 & 1 & 1 \\
    1 & -1 & 1 & -1 \\
    1 & 1 & -1 & -1 \\
    1 & -1 & -1 & 1 
\end{bmatrix}\vec{F}\]

\[\vec{w} = (W_2^{-1}\vec{F}), \]
where $W_2$ is the second-order Walsh matrix. 

Expression 8 can also be extended to three variables and after decomposition, the following canonical expression is derived:
\begin{center}
    \scalebox{0.79}{$a^{f(x_1,x_2,x_3)} = [[((a^{w_a}g^{x_3}a^{w_b}g^{x_3})g^{x_2})((a^{w_c}g^{x_3}a^{w_d}g^{x_3})g^{x_2})]g^{x_1}][[((a^{w_e}g^{x_3}a^{w_f}g^{x_3})g^{x_2})((a^{w_g}g^{x_3}a^{w_h}g^{x_3})g^{x_2})]g^{x_1}]$} \\ $= a^{w_a}g^{x_3}a^{w_b}g^{x_2 + x_3}a^{w_c}g^{x_3}a^{w_d}g^{x_1+x_2 + x_3}a^{w_e}g^{x_3}a^{w_f}g^{x_2 + x_3}a^{w_g}g^{x_3}a^{w_h}g^{x_1 + x_2+x_3}$. 
\end{center}

In general, \[F(x_1,x_2, \ldots x_n, x_{n+1}) = F_a(x_2, \ldots x_{n+1})g^{x_1}F_b(x_2, \ldots x_{n+1})g^{x_1} \text{ and }\vec{w} = (W_{n+1}^{-1}\vec{F}).\]

\end{document}